\documentclass[a4paper,11pt]{article}
\usepackage{jheppub} 
\usepackage{lineno}
\usepackage{array}

\newcommand{\Na}{$^{11}{\rm C^*} \rightarrow  n +  ^{10}{\rm C}$\ }
\newcommand{\Nb}{$^{11}{\rm C^*} \rightarrow   n + \gamma +  ^{10}{\rm C}$\ }
\newcommand{\NNa}{$^{10}{\rm C^*} \rightarrow   n +  ^{9}{\rm C}$\ }
\newcommand{\NNb}{$^{10}{\rm C^*} \rightarrow   n + p +  ^{8}{\rm B}$\ }

\newcommand{\Ninv}{$ n \rightarrow inv$\ }
\newcommand{\NNinv}{$ n n \rightarrow inv$\ }

\title{\boldmath JUNO Sensitivity to Invisible Decay Modes of Neutrons}

\collaboration{JUNO collaboration}
\author[6,5]{Angel Abusleme}
\author[45]{Thomas Adam}
\author[48]{Kai Adamowicz}
\author[66]{Shakeel Ahmad}
\author[66]{Rizwan Ahmed}
\author[55]{Sebastiano Aiello}
\author[21]{Fengpeng An}
\author[23]{Qi An}
\author[55]{Giuseppe Andronico}
\author[67]{Nikolay Anfimov}
% ...
\affiliation[5]{Millennium Institute for SubAtomic Physics at the High-energy Frontier (SAPHIR), ANID, Chile}
\affiliation[6]{Pontificia Universidad Cat\'{o}lica de Chile, Santiago, Chile}
% ...
\affiliation[21]{Sun Yat-Sen University, Guangzhou, China}
% ...
\affiliation[23]{University of Science and Technology of China, Hefei, China}
% ...
\affiliation[45]{IPHC, Universit\'{e} de Strasbourg, CNRS/IN2P3, F-67037 Strasbourg, France}
% ...
\affiliation[48]{III. Physikalisches Institut B, RWTH Aachen University, Aachen, Germany}
% ...
\affiliation[55]{INFN Catania and Dipartimento di Fisica e Astronomia dell Universit\`{a} di Catania, Catania, Italy}
% ...
\affiliation[66]{Pakistan Institute of Nuclear Science and Technology, Islamabad, Pakistan}
\affiliation[67]{Joint Institute for Nuclear Research, Dubna, Russia}
% ...

\emailAdd{Juno\_pub\_comm@juno.ihep.ac.cn}

\abstract{We explore the decay of bound neutrons into invisible particles (e.g., $n\rightarrow 3 \nu$ or $nn \rightarrow 2 \nu$) in the JUNO liquid scintillator detector, which do not produce an observable signal. The invisible decay includes two decay modes: $ n \rightarrow { inv} $ and $ nn \rightarrow { inv} $. The invisible decays of $s$-shell neutrons in $^{12}{\rm C}$ will leave a highly excited residual nucleus. Subsequently, some de-excitation modes of the excited residual nuclei can produce a time- and space-correlated triple coincidence signal in the JUNO detector. Based on a full Monte Carlo simulation informed with the latest available data,  we estimate all backgrounds, including inverse beta decay events of the reactor antineutrino $\bar{\nu}_e$, natural radioactivity, cosmogenic isotopes and neutral current interactions of atmospheric neutrinos. Pulse shape discrimination and multivariate analysis techniques are employed to further suppress backgrounds. With two years of exposure, JUNO is expected to give an order of magnitude improvement compared to the current best limits. After 10 years of data taking, the JUNO expected sensitivities at a 90\% confidence level are $\tau/B( n \rightarrow { inv} ) > 5.0 \times 10^{31} \, {\rm yr}$ and $\tau/B( nn \rightarrow { inv} ) > 1.4  \times 10^{32} \, {\rm yr}$.}

\begin{document}
\maketitle
\flushbottom

\section{Introduction}\label{sec:intro}
The conservation of the baryon number $B$ is an accidental symmetry in the Standard Model of particle physics, and no fundamental symmetry guarantees the proton's stability. Baryon number violation is one of three basic ingredients to generate the cosmological matter-antimatter asymmetry from an initially symmetrical universe \cite{Sakharov:1967dj}. On the other hand, baryon number $B$ is necessarily violated, and the proton must decay in the Grand Unified Theories (GUTs) \cite{Nath:2006ut}, which can unify the strong, weak, and electromagnetic interactions into a single underlying force. These GUTs have motivated a long history of experiments searching for proton decay \cite{ParticleDataGroup:1994kdp}. However, no experimental evidence to date for proton decay or $B$-violating neutron decay has been found \cite{ParticleDataGroup:2022pth}. Discovering nucleon decay now remains a key signature of GUTs. Super-Kamiokande with Gd-loaded water can improve its background rejection capability in the nucleon decay searches \cite{Super-Kamiokande:2021the}. The new generation of underground experiments such as JUNO \cite{JUNO_YB,JUNO_PPNP,JUNO_CDR}, Hyper-Kamiokande \cite{Hyper-K} and DUNE \cite{DUNE} with tens or even hundreds of kiloton target masses and different detection technologies, will continue to search for the nucleon decay and test GUTs.

JUNO is a 20 kton multipurpose underground liquid scintillator (LS) detector under construction in China, with a 650-meter rock overburden (1800 m.w.e.) for shielding against cosmic rays \cite{JUNO_YB, JUNO_PPNP, JUNO_CDR}. The LS detectors have distinct advantages in the search for some nucleon decay modes, such as $p\rightarrow \bar{\nu} K^+$ \cite{LENA,KamLAND_vK,JUNO:2022qgr} and the neutron invisible decays \cite{Kamyshkov:2002wp,KamLAND_invisible}. The proton decay mode $p\rightarrow \bar{\nu} K^+$ is one of the two dominant decay modes predicted by a majority of GUTs \cite{Babu:2013jba}. The JUNO expected sensitivity will reach $\tau/B(p\rightarrow \bar{\nu} K^+) > 9.6 \times 10^{33}$~yr with 10 years of data taking \cite{JUNO:2022qgr}, which is higher than the current best limit of $5.9 \times 10^{33}$ yr from the Super-Kamiokande experiment \cite{Abe:2014mwa}. The neutron invisible decay has two modes: \Ninv and $nn \rightarrow inv$. Invisible modes are dominant in some new physics models~\cite{Mohapatra:2002ug, Barducci:2018rlx,Girmohanta:2019fsx,Girmohanta:2020eav}. The SNO+ experiment sets the current best limit for single neutron disappearance at $\tau/B(n \rightarrow \textit{inv}) > 9.0 \times 10^{29}$ years (90\% C.L.), while for the $nn \rightarrow inv$ mode, the best limit is provided by the KamLAND experiment with $\tau/B(nn \rightarrow inv) > 1.4 \times 10^{30}$ years (90\% C.L.)~\cite{Kamyshkov:2002wp,SNO:2022trz, KamLAND_invisible}.

In this paper, the JUNO potential for neutron invisible decays is investigated. Sec.~\ref{sec:JUNO_detector} briefly introduces the JUNO detector and its expected performance. In Sec.~\ref{sec:simulation}, we describe the Monte Carlo (MC) simulation of neutron invisible decays and all background sources. In Sec.~\ref{sec:es}, some basic event selection criteria are developed to discriminate the neutron invisible decays from the backgrounds. In Sec.~\ref{sec:bkg_estimation}, we estimate all possible backgrounds in detail and use the pulse shape discrimination and multivariate analysis techniques to further suppress them. Sec.~\ref{sec:sen} presents the expected JUNO sensitivities to the neutron invisible decays. Finally, a conclusion is given in Sec.~\ref{sec:conclusion}.

\section{JUNO detector}\label{sec:JUNO_detector}

\begin{figure}[]
	\centering
	\includegraphics[width=0.8\textwidth]{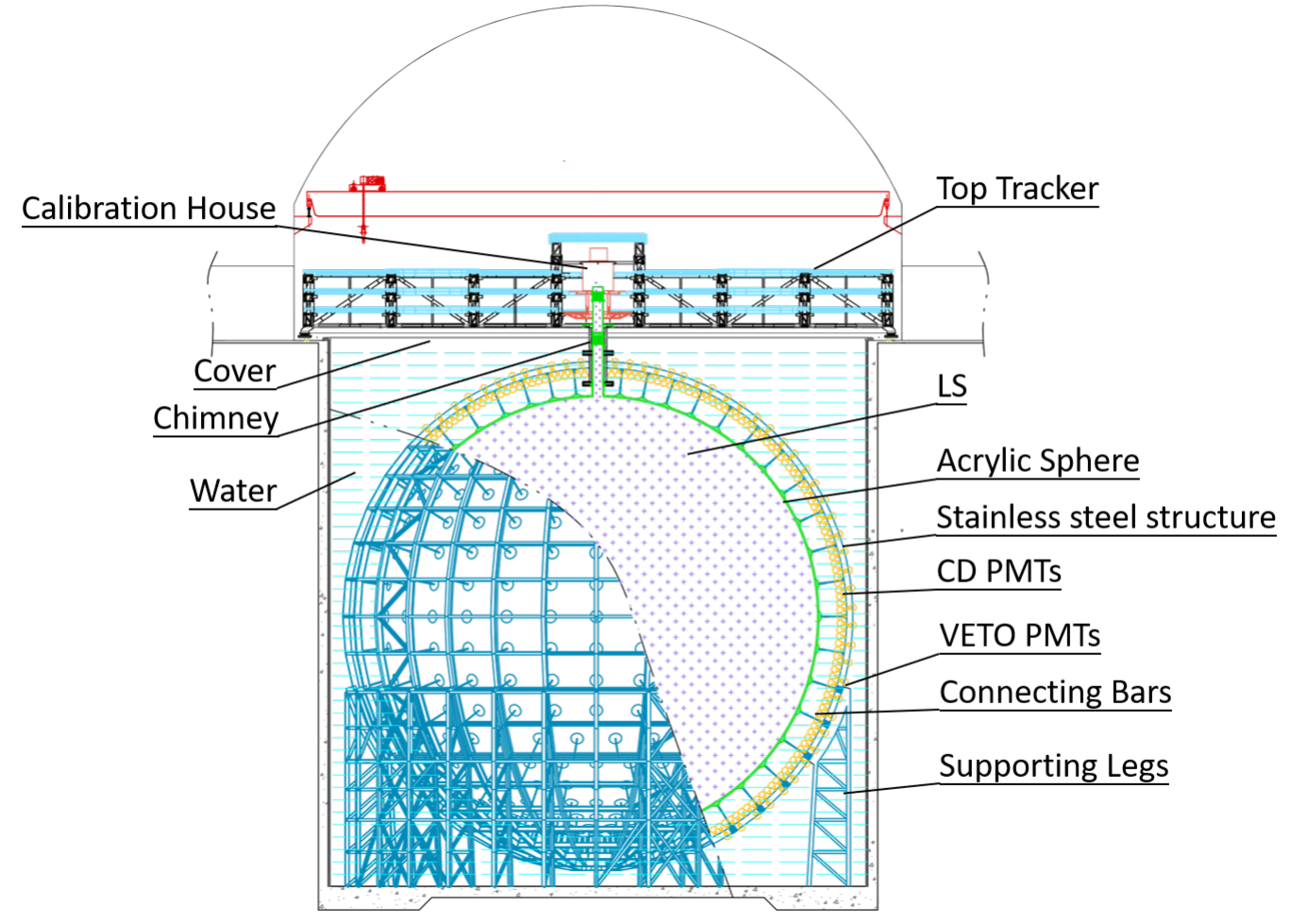}
	\caption{\label{fig:detector} Schematic of the JUNO detector.}
\end{figure}

As a multipurpose neutrino observatory, JUNO comprises a central detector (CD), veto detector, and calibration system \cite{JUNO_PPNP,JUNO_CDR} as shown in Fig.~\ref{fig:detector}. The CD holds 20~kton of LS \cite{JUNO:2020bcl} filled in an acrylic shell with an inner diameter of 35.4 m, which is immersed in a cylindrical water pool (WP) with both diameter and height of 43.5 m. There are 17,612 high quantum efficiency 20-inch PMTs (LPMTs) and 25,600 3-inch PMTs (SPMTs), yielding an integral 77.9\% photo-cathode coverage, closely packed around the LS ball, aiming to achieve an energy resolution of $\leq$ 3\% at 1 MeV. Two kinds of LPMTs, MCP-PMT from NNVT and dynode PMT from Hamamatsu, are used \cite{JUNO:2022hlz}. The veto detector, optically decoupled from the CD, is designed to tag cosmic muons with high efficiency and to precisely track performance for the purpose of background reduction. The veto system includes a water Cherenkov detector(WCD) and a top tracker (TT) surrounding the CD to shield the neutrons and the natural radioactivity from the rock. The water Cherenkov detector is made of 35 kton of ultrapure water with a radon concentration below 0.2 Bq/m$^3$ \cite{JUNO_PPNP}, which is supplied and maintained by a circulation system. The Cherenkov light is detected by 2400 20-inch PMTs, and its muon detection efficiency is expected to reach 99.5$\%$ \cite{JUNO_PPNP}. The top tracker is made of reused plastic scintillators from the OPERA experiment, and it covers half of the water pool on top with a 3-layer configuration for the muon track angular reconstruction precision of 0.20$^{\circ}$ \cite{JUNO:2023cbw}. The calibration system includes the deployment of sources along the central axis from an automated calibration unit (ACU). It also consists of a cable loop system which allows for position these sources to off-axis positions on a plane. Additionally, there is a remotely operated vehicle and a guide tube system outside of the acrylic sphere to study boundary effects~\cite{JUNO:2020xtj,Zhu:2022ovo}.
 
\section{Simulation} \label{sec:simulation}

To understand the detector signature of neutron invisible decays and discriminate them from the backgrounds in the JUNO detector, a full MC simulation has been performed with the JUNO simulation framework~\cite{Lin:2022htc}. The simulation accounts for the response of the detector and electronics, as well as the performance of the energy and vertex reconstructions. The detector simulation employs GEANT4~\cite{GEANT4}. The electronics simulation~\cite{Jetter:2012xp} incorporates factors such as dark noise, transit time spread, and afterpulses. To build the hit time and charge information of each readout signal, we adopt the deconvolution method in the step of waveform reconstruction \cite{Huang:2017abb}. Furthermore, the maximum likelihood estimation strategy from Ref.~\cite{Huang:2022zum}, combining charge and time, is used for energy and vertex reconstruction. All background sources are described and simulated in this section, including the inverse beta decay (IBD) events from the reactor antineutrino $\bar{\nu}_e$, natural radioactivity, cosmogenic isotopes and fast neutrons from the cosmic-ray muons, and atmospheric neutrino events.

\subsection{Neutron invisible decays} \label{subsec:sig_features}

The JUNO LS includes about $88\%$ $^{12}{\rm C}$ and $12\%$ $^{1}{\rm H}$ \cite{JUNO_YB}. The invisible decays of neutrons from the $s$-shell in $^{12}{\rm C}$ will lead to a highly excited residual nucleus. Then the excited nucleus can emit secondary particles ($p, n, \alpha, \gamma$) and leave a daughter nucleus. In Ref.~\cite{Kamyshkov:2002wp}, Kamyshkov and Kolbe have analyzed one and two neutron invisible decays from $^{12}{\rm C}$ based on the statistical model code SMOKER \cite{Cowan:1991zz}. It has been found that some de-excitation modes of the excited nucleus can give time-, energy-, and space-correlated signals in the LS detector. Here we consider the following four de-excitation modes \cite{Kamyshkov:2002wp,Cowan:1991zz}:
\begin{eqnarray}
 ^{11}{\rm C^*}  \rightarrow &  n + ^{10}{\rm C}           \;\;\;\;   &  (B_{n1} = 3.0\%),  \label{M1} \\
 ^{11}{\rm C^*}  \rightarrow &  n + \gamma  + ^{10}{\rm C} \;\;\;\;   &  (B_{n2} = 2.8\%),  \label{M2}\\
 ^{10}{\rm C^*}  \rightarrow &  n + ^{9}{\rm C}                 \;\;\;\;   &  (B_{nn1} = 6.2\%), \label{M3}\\
 ^{10}{\rm C^*}  \rightarrow &  n + p + ^{8}{\rm B}             \;\;\;\;   &  (B_{nn2} = 6.0\%), \label{M4}
\end{eqnarray}
where the daughter nuclei $^{10}{\rm C }$($\beta^+$,  19.3 s, 3.65 MeV), $^{9}{\rm C}$($\beta^+$, 0.127 s, 16.5 MeV), and $^{8}{\rm B}$($\beta^+ \alpha$, 0.770 s, 18.0 MeV) are radioactive. The corresponding decay mode, half-life, and energy release have been indicated in parentheses. The de-excitation modes in Eqs.~(\ref{M1}) and (\ref{M2}) for the single neutron invisible decay $n \rightarrow inv$ have branching ratios of $B_{n1}=3.0\%$ and $B_{n2}=2.8\%$, respectively. For the two neutron invisible decays $nn \rightarrow inv$, the de-excitation modes in Eqs.~(\ref{M3}) and (\ref{M4}) exhibit the branching ratios of $B_{nn1} =6.2\%$ and $B_{nn2} =6.0\%$.

Note that the above four de-excitation modes feature a triple coincidence signal in the LS detector \cite{Kamyshkov:2002wp}. The first signal comes from neutron elastic and inelastic scatterings with free protons and $^{12}{\rm C}$. The de-excitation products $\gamma$ and $p$ can also contribute to this first signal of the triple coincidence event. The neutron will quickly slow down and be thermalized after many collisions. In the LS, these thermalized neutrons are captured by a free proton $\sim 220 \, \mu{\rm s}$ later and give a second signal of the 2.2 MeV $\gamma$ ray. The third signal arises from the $\beta^+$ decay of the daughter nuclei $^{10}{\rm C}$, $^{9}{\rm C}$, and $^{8}{\rm B}$. The strong time-, energy-, and space-correlation between the three signals can be exploited to significantly suppress backgrounds.

We have made an event generator and used it to generate 0.5 million events for each de-excitation mode in Eqs.~(\ref{M1})-(\ref{M4}). Here, the neutron kinetic energy distributions in Figs.~4 and 6 of Ref.~\cite{Kamyshkov:2002wp} have been used. The emitted $\gamma$ in Eq.~(\ref{M2}) is dominated by a strong monoenergetic line at 3.35 MeV, which is produced from the decay of the first excited $2^+$ state of $^{10}$C~\cite{Kamyshkov:2002wp}. The ejected proton from Eq.~(\ref{M4}) is monoenergetic with an energy of 0.922 MeV, which corresponds to the de-excitation of the first excited state of $^9$C \cite{Kamyshkov:2002wp}. The kinetic energies of the daughter nuclei $^{10}{\rm C}$, $^{9}$C, and $^{8}$B have been set to zero. Then we simulate these events from \Ninv and $nn \rightarrow inv$ and obtain the energy, spatial, and time interval distributions of these triple coincidence signals. In this work, $E_{i}$ ($R_{i}$) is used to denote the reconstructed energy (radial position) of the $i^{th}$ signal with $i = 1, 2, 3$. $\Delta R_{ij}$ and $\Delta T_{ij}$ ($i,j = 1,2,3; i < j$) describe the distance and time interval between the $i^{th}$ signal and the $j^{th}$ signal, respectively. It is worth noting that some invisible decay events are unable to exhibit the triple coincidence feature due to energy leakage, which is the loss of energy resulting from its deposition in non-active volumes. Conversely, some events yield a coincidence of more than three signals, for example, if two neutrons are captured or if radioactive isotopes are induced by energetic neutrons. The detection efficiencies for the four de-excitation modes in Eqs.~(\ref{M1})-(\ref{M4}) are 91.7\%, 95.6\%, 90.4\%, and 90.5\%. In order to better reflect the physical characteristics, we refer to the first, second, and third signals as prompt, delayed, and decay signals, respectively.   

\begin{figure}[!t]
	\centering
	\includegraphics[width=0.325\textwidth]{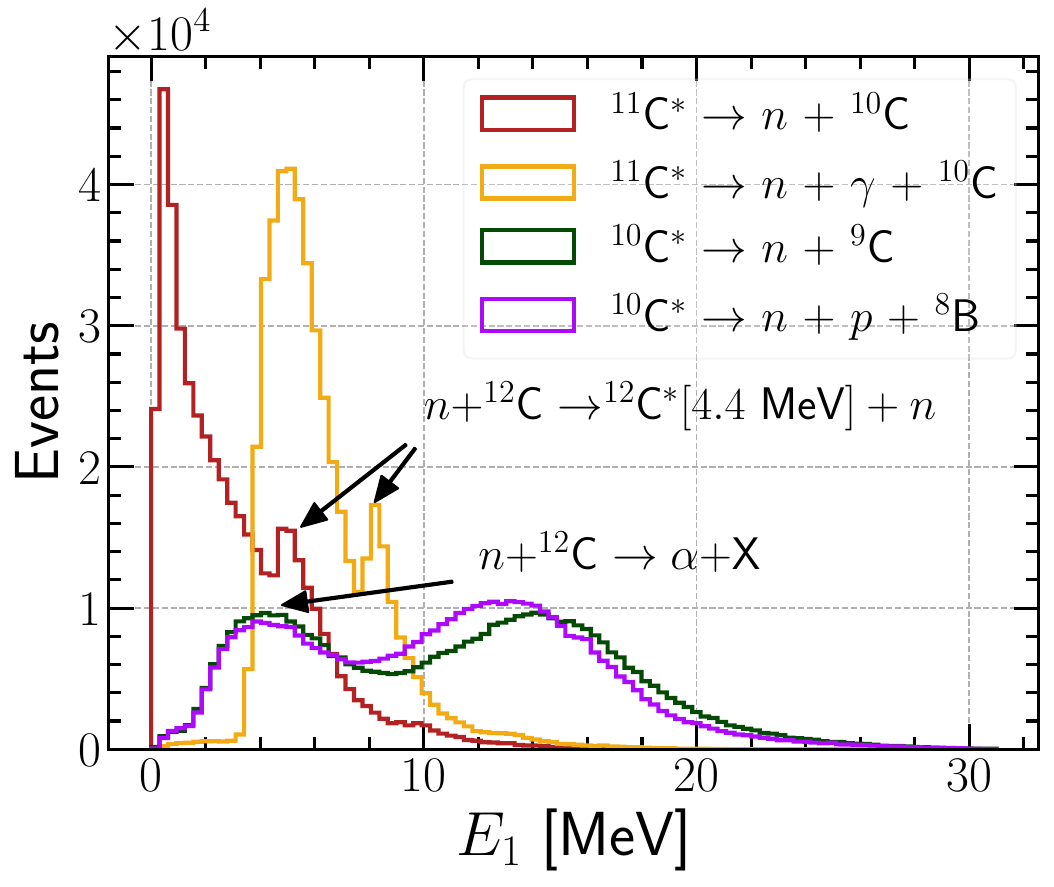}
	\includegraphics[width=0.325\textwidth]{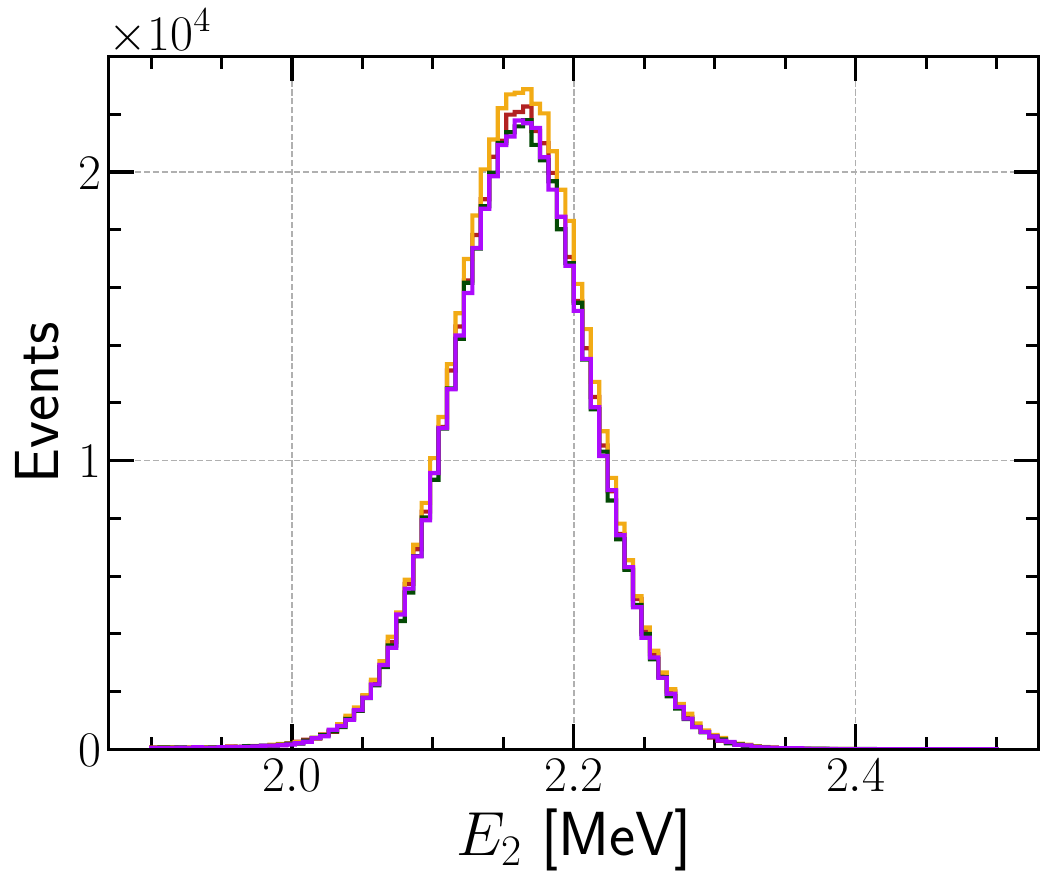}
	\includegraphics[width=0.325\textwidth]{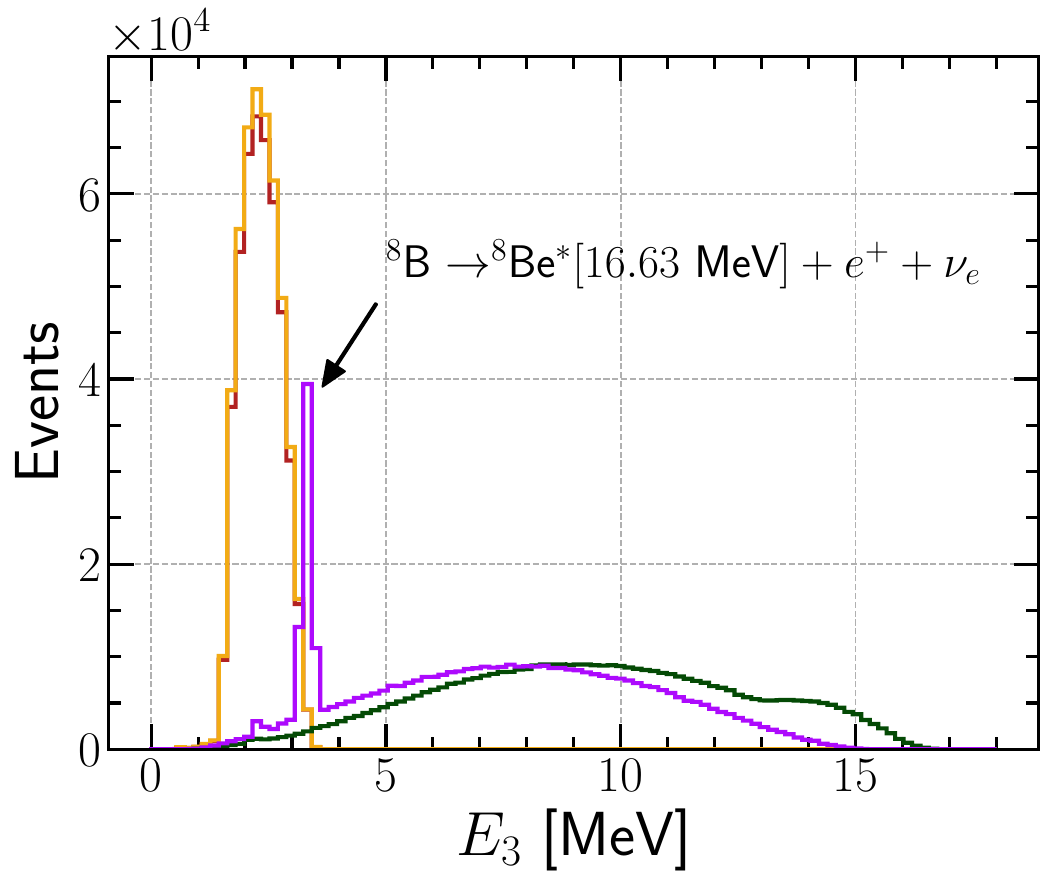}
	\includegraphics[width=0.325\textwidth]{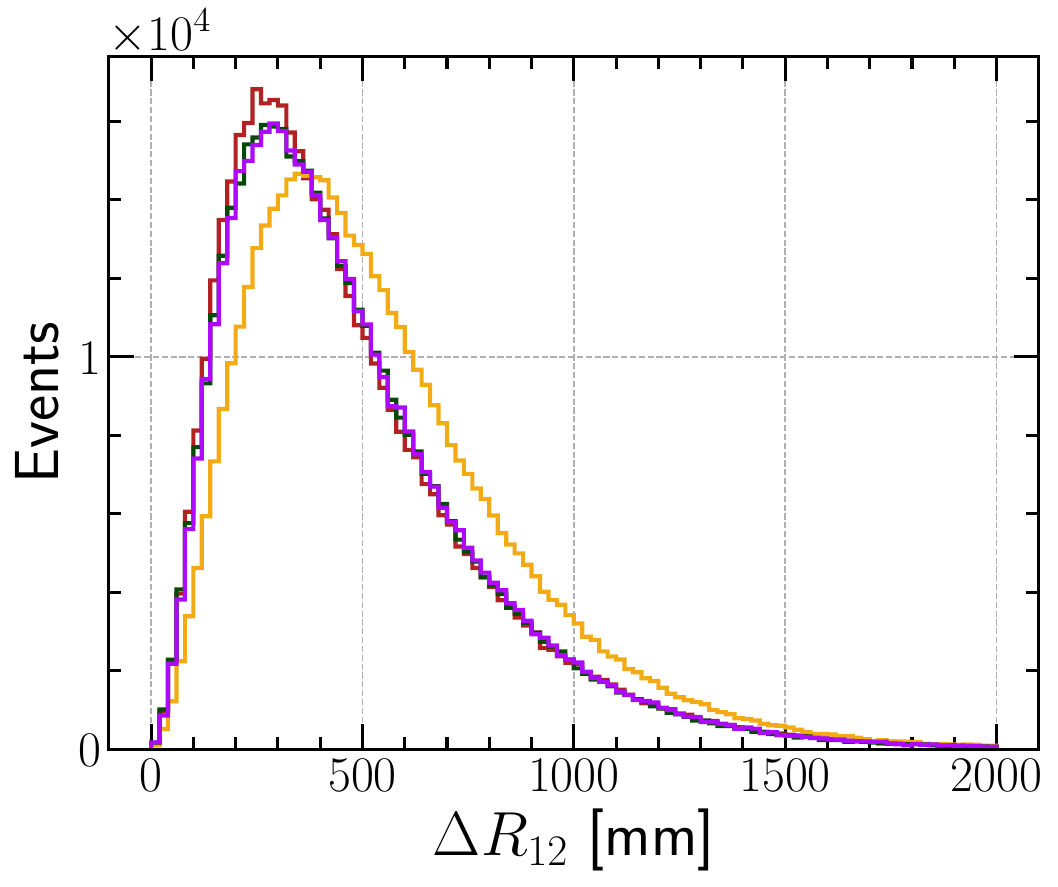}
	\includegraphics[width=0.325\textwidth]{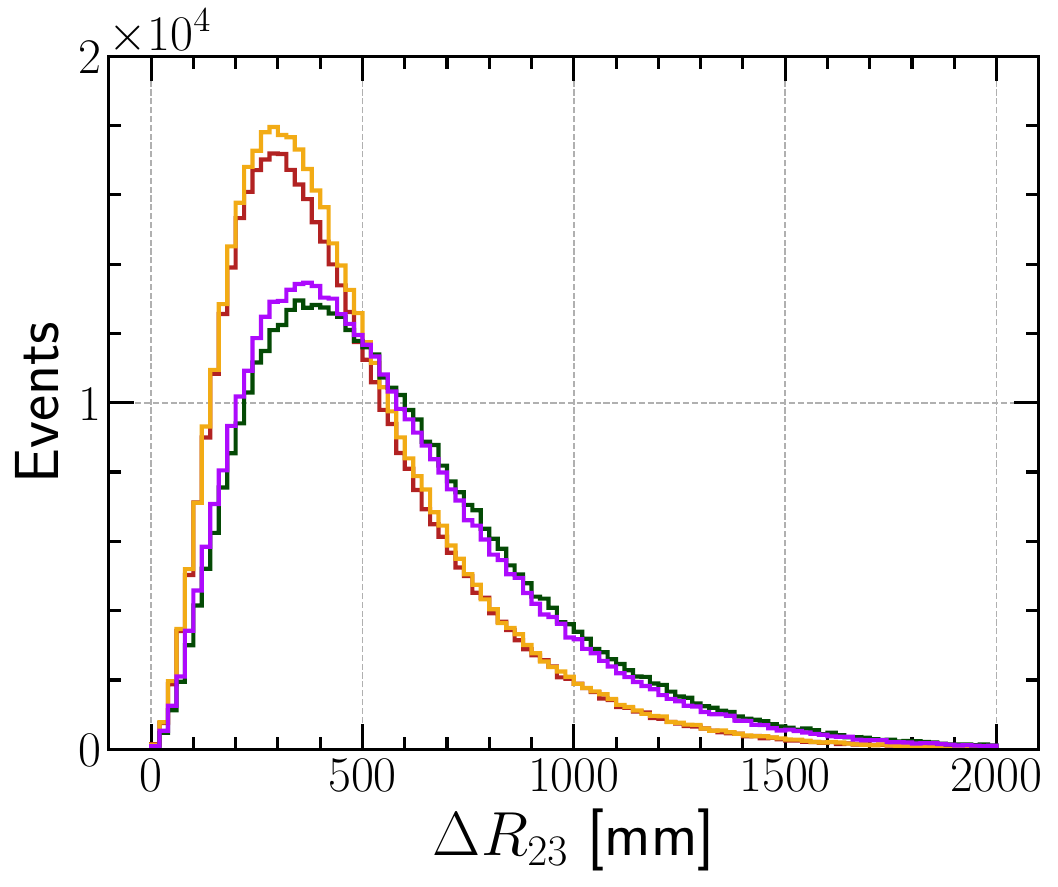}
	\includegraphics[width=0.325\textwidth]{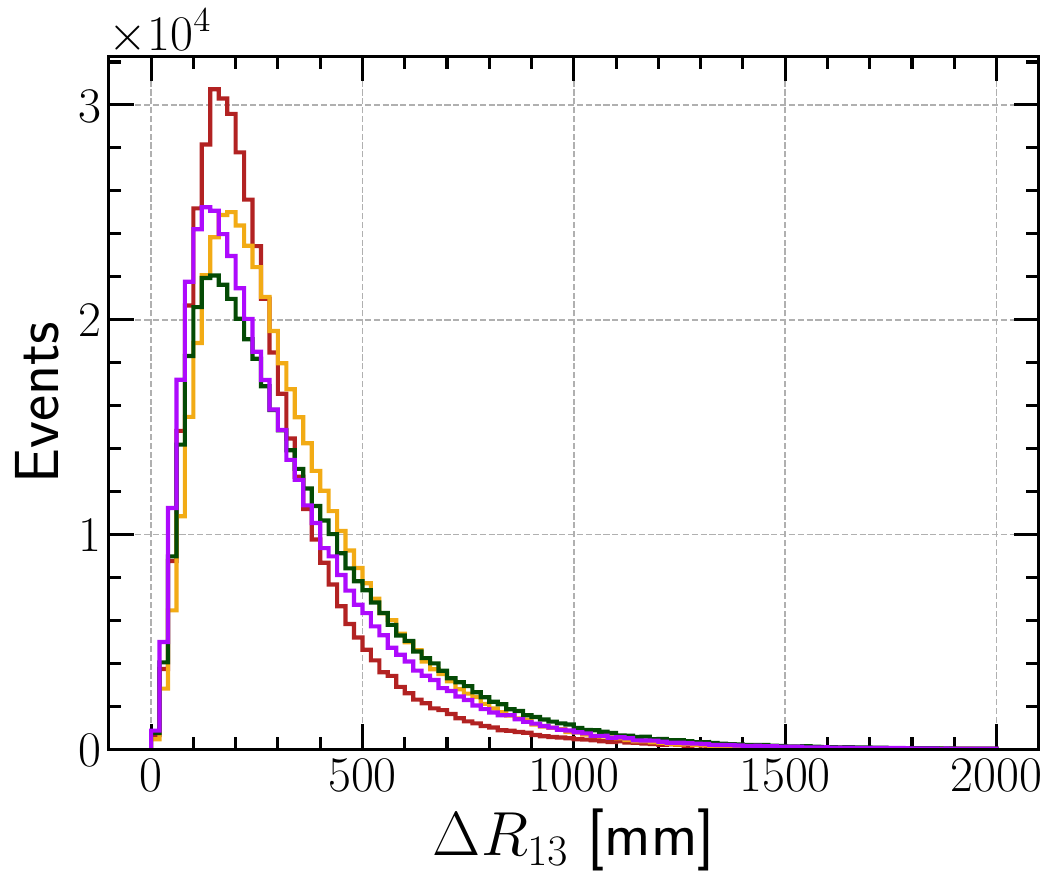}
    \caption{The $E_1$, $E_2$, $E_3$, $\Delta R_{12}$, $\Delta R_{13}$, and $\Delta R_{23}$ distributions of four de-excitation modes.}
	\label{fig:Singal_characteristic}
\end{figure}

The distributions of $E_{i}$ and $\Delta R_{ij}$ from the four de-excitation modes in Eqs.~(\ref{M1})-(\ref{M4}) are displayed in Fig.~\ref{fig:Singal_characteristic}. All prompt energy spectra exhibit two peaks because of the neutron elastic and inelastic scattering processes. For $^{11}{\rm C^*}  \rightarrow  n + ^{10}{\rm C}$ and $^{11}{\rm C^*}  \rightarrow  n + \gamma + ^{10}{\rm C}$, the higher energy peak of the $E_{1}$ distribution comes from the neutron inelastic scattering with $^{12}$C, which emits a $\gamma$ ray of 4.4~MeV. Due to the larger initial kinetic energy, neutrons from $ ^{10}{\rm C^*}  \rightarrow n + ^{9}{\rm C} $ and $^{10}{\rm C^*}  \rightarrow   n + p + ^{8}{\rm B}$  can produce an $\alpha$ particle through the inelastic scattering $n + ^{12}{\rm C} \to \alpha + {\rm X}$, which corresponds to the lower energy peaks in the top-left panel of Fig.~\ref{fig:Singal_characteristic}. For the decay signal, there is a very narrow peak around 3.3 MeV caused by the $\beta^+$ decay of $^{8}{\rm B} \rightarrow ^8{{\rm Be}^*}(16.63~ {\rm MeV}) + e^{+} + \nu_{e}$. This occurs because the excited state of $^8{{\rm Be}^*}$~(16.63 MeV) will subsequently decay into two $\alpha$ particles, which has a large quenching factor in the LS. The spatial correlations of the triple coincidence signal show that the distances $\Delta R_{12}$, $\Delta R_{13}$ and $\Delta R_{23}$ are mostly less than 1.5 m. It is worthwhile to stress that all distributions of $E_{i}$, $\Delta R_{ij}$, and $\Delta T_{ij}$ from the MC simulation are in agreement with the expectations from the National Nuclear Data Center~\cite{nndc} and the International Atomic Energy Agency Nuclear Data Services~\cite{iaea-nds}.

\subsection{Background sources}\label{subsec:background_sources}

As discussed in the preceding subsection, the neutron invisible decays can generate a triple coincidence signal in the LS detector, which has been exploited by the KamLAND experiment to perform this search \cite{KamLAND_invisible}. The dominant background stems from the doubly correlated IBD event ($p + \bar{\nu}_e \rightarrow e^{+} + n$) followed by an uncorrelated natural radioactive decay. Unlike KamLAND, JUNO has a larger target mass and shallower rock overburden, potentially resulting in a wider variety of background sources. Therefore, we thoroughly investigate and simulate the various background sources, including IBD events, natural radioactivity, cosmogenic isotopes, fast neutrons (FN), and atmospheric neutrino (Atm-$\nu$) events. In Sec.~\ref{sec:bkg_estimation}, we will assess the JUNO background for the \Ninv and \NNinv searches by leveraging these MC samples of background sources.  

In JUNO \cite{JUNO_YB, JUNO_PPNP}, the reactor antineutrinos are detected by the IBD reaction via the prompt-delayed coincidence signal. The kinetic energy deposited by the positron via ionization, together with its subsequent annihilation into typically two 0.511 MeV $\gamma$s, forms a prompt signal. The impinging neutrino transfers most of its energy to the positron. This establishes a strong correlation between the reconstructed energy of the positron and the energy of the antineutrino $\bar{\nu}_{e}$, which is a crucial parameter for measuring neutrino oscillations. The neutron is captured in an average time $\sim 220$ $\mu s$, and the corresponding photon emission forms a delayed signal. Neutron captures predominantly occur on hydrogen ($\sim 99\%$), resulting in the release of a single 2.2 MeV $\gamma$ ray, while neutron captures on carbon ($\sim 1 \%$) yield a gamma-ray signal with a total energy of 4.9 MeV, albeit very infrequently. The expected average IBD rate in JUNO is 57.4 /day~\cite{JUNO:2022mxj}. The accidental coincidence between an IBD event and an uncorrelated single event significantly contributes to the background.

\begin{figure}[!t]
	\centering
	\includegraphics[width=0.6\textwidth]{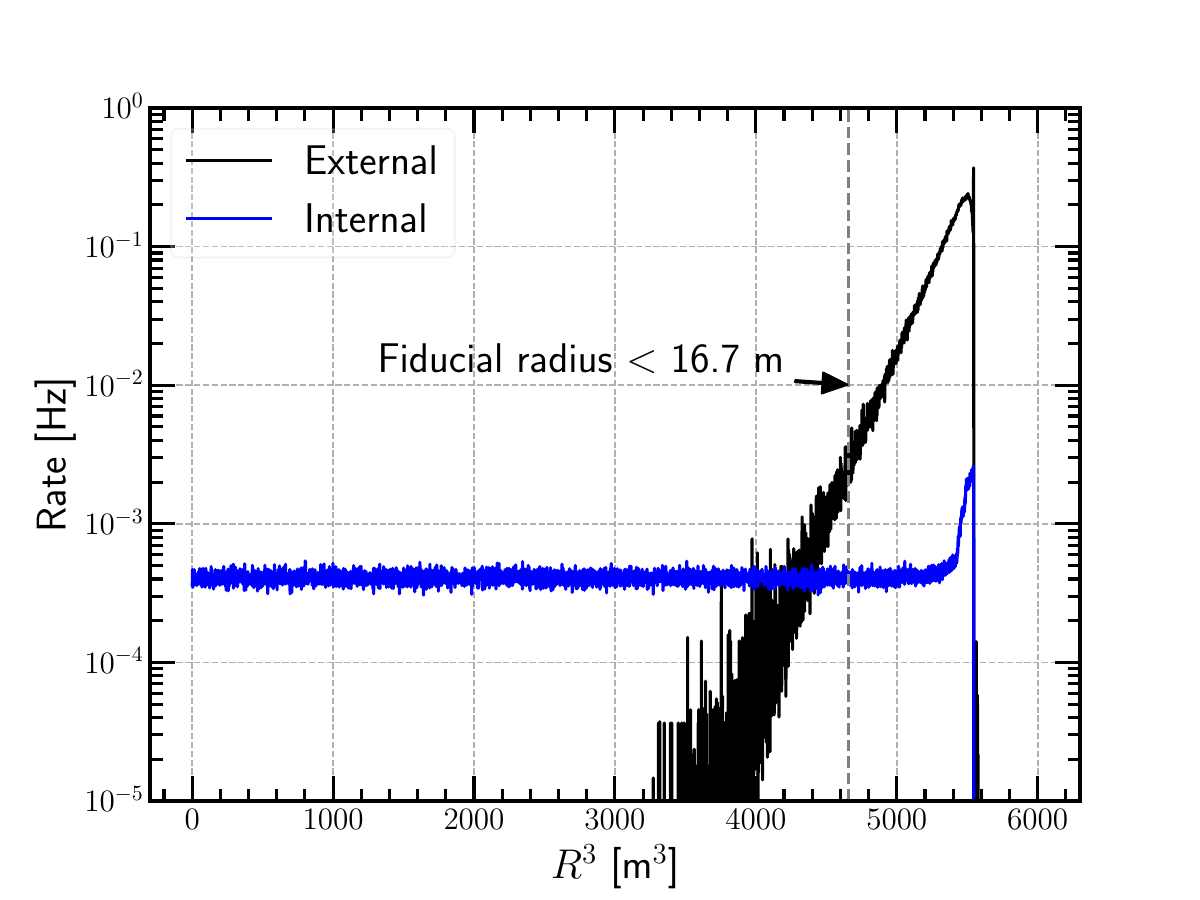}
	\caption{\label{fig:radioactivity_fv}The internal and external natural radioactivity rates as a function of radius. The dashed line represents the fiducial radius.}
\end{figure}

Natural radioactivity is found in all materials and can only be reduced by strict requirements for material screening and environmental control. It can be separated into internal radioactivity originating from the LS itself and external radioactivity from other parts of the JUNO detector, with the predominant external background originating in the PMT glass. The internal radioactivity primarily comes from radio-nuclides in the U/Th chains, with the assumed concentration of U/Th being $10^{-6}$ ppb~\cite{JUNO:2021kxb}. They contribute to the deposited energy regardless of the type of emitted particles ($\alpha$, $\beta$, and $\gamma$), as the energy is directly released into the LS, where these radionuclides are uniformly distributed. To reduce the impact of radioactivity, JUNO has implemented strict background control strategies, enabling the rate of natural radioactivity to be limited to 10~Hz. Ref.~\cite{JUNO:2021kxb} provides a comprehensive summary of the approach taken by JUNO to achieve this goal and presents the internal and external radioactivity results from the detector simulation. The external radioactivity can be suppressed significantly with a fiducial volume cut of $R_{i}<$ 16.7 m and an energy cut of 0.7--30 MeV as shown in Fig.~\ref{fig:radioactivity_fv}. In this case, the internal and external radioactivity rates are around 1.98 Hz and 0.48 Hz. Meanwhile, the $\alpha$ particle from $^{238}$U, $^{232}$Th, $^{210}$Po radioactive decay chains can interact with $^{13}$C in LS, and the interaction of $^{13}$C($\alpha$, n)$^{16}$O may produce the prompt-delayed event. The expected rate of such events can be determined based on the concentrations~\cite{JUNO:2021kxb} and neutron yields~\cite{Zhao:2013mba}.

Cosmogenic radioactive isotopes and neutrons can be produced in the LS through the spallation of cosmic-ray muons. Approximately 3.94 Hz of muons will pass through the JUNO LS with an average energy of 207~GeV~\cite{JUNO_PPNP}. MUSIC \cite{Kudryavtsev:2008qh} is employed to track muons traversing the rock to the underground experimental hall, based on the local geological map. Among the generated cosmogenic isotopes, the long-lived isotopes $^{9}$Li and $^{8}$He can undergo $\beta^-n$ decay, which can mimic the prompt and delayed signals of neutron invisible decays. Other long-lived isotopes will give a single signal similar to natural radioactivity, which predominantly comes from $^{11}$C, $^{12}$B, $^{6}$He, $^{10}$C, and $^{8}$Li. To address the observed discrepancies between the GEANT4~\cite{GEANT4} simulation and actual isotope yield data, the cosmogenic isotope rates are adjusted based on the experimental results from both KamLAND~\cite{KamLAND:2009zwo} and Borexino \cite{Borexino:2008fkj}. The event rates for $^9$Li and $^8$He are determined to be 127 per day and 40 per day, while the event rates of $^{11}$C, $^{12}$B, $^{6}$He, $^{10}$C and $^{8}$Li  are 50020, 2478, 2373, 953 and 705 per day, respectively. To enhance the precision of cosmogenic isotope results and study the muon veto strategy, we simulated a decade of muon data without including the photon production and propagation processes.

Fast neutrons, originating from untagged cosmic muons and coinciding with an uncorrelated single event, is another background that needs to be considered. The tagged efficiencies of muons passing through the LS and only the water buffer are almost 100\% and 99.8\% \cite{JUNO_YB}, respectively. The untagged muon is also called a rock muon, including the corner clipping muons and the muon whose track length in water is less than 50 cm. Fast neutrons associated with the untagged muons may enter the CD and produce a correlated signal, which can mimic the prompt and delayed signals, whose rate is $\sim$ 0.4 /day. Note that the fast neutrons are primarily concentrated around the CD edge. Accordingly, the fiducial volume cut of $R_i <$ 16.7 m lowers their rate to about ~0.08 /day.

Atmospheric neutrino (Atm-$\nu$) charged current (CC) and neutral current (NC) interactions can also contribute to the background of neutron invisible decays. Atm-$\nu$ CC events usually have an energy larger than the typical nuclear de-excitation energy \cite{Kamyshkov:2002wp}. For the prompt signal energy range from 0.7 to 30.0 MeV, the IBD-like event rate from the Atm-$\nu$ CC interaction has been estimated by Ref.~\cite{JUNO:2022lpc}. It is evident that the IBD-like event rate is significantly lower than the JUNO reactor IBD rate. Here, we have found that the triple coincident events from Atm-$\nu$ CC reactions are negligible. 
An assessment of the expected CC and NC rates from atmospheric neutrinos below 100~MeV \cite{Battistoni:2005pd} resulted in a negligible contribution. Hence, our analysis primarily concentrates on NC background from atmospheric neutrinos with energies higher than 100 MeV. As listed in Table 2 of Ref.~\cite{KL30} and Table I of Ref. \cite{Mollenberg:2014pwa}, the following three Atm-$\nu$ NC processes may directly form a triple coincidence signal:
\begin{eqnarray}
\nu/\bar{\nu} + ^{12}{\rm C} & \rightarrow & \nu/\bar{\nu} + n  +
^{11}{\rm C}\;,  \label{C11} \\
\nu/\bar{\nu} + ^{12}{\rm C} & \rightarrow & \nu/\bar{\nu} + 2n +
^{10}{\rm C}\;,  \label{C10}\\
\nu/\bar{\nu} + ^{12}{\rm C} & \rightarrow & \nu/\bar{\nu} + 3 p + n
+ ^{8}{\rm Li}\;,\label{Li8}
\end{eqnarray}
where the daughter nuclei $^{11}{\rm C}$($\beta^+$, 20.36 min, 1.98 MeV), $^{10}{\rm C}$($\beta^+$,  19.3 s, 3.65 MeV) and $^{8}{\rm Li}$($\beta^- \alpha$, 0.839 s, 16.0 MeV) are radioactive. The reaction in Eq.~(\ref{C10}) can produce only one neutron capture signal if one of two neutrons loses energy at the CD edge or disappears through the inelastic scattering with $^{12}{\rm C}$. In addition to the three processes, some Atm-$\nu$ NC reactions can also form the triple signal indirectly through accidental coincidences between the prompt-delayed signal and an uncorrelated single event. The MC simulation indicates that over 90\% of all prompt-delayed signals with $E_1 < 30$ MeV are generated by four dominant NC interactions:
\begin{eqnarray}
\nu/\bar{\nu} + ^{12}{\rm C} & \rightarrow & \nu/\bar{\nu} + n + p +^{10}{\rm B}\;,  \label{B10} \\
\nu/\bar{\nu} + ^{12}{\rm C} & \rightarrow & \nu/\bar{\nu} + n + p + \alpha + ^{6}{\rm Li}\;,  \label{Li6}\\
\nu/\bar{\nu} + ^{12}{\rm C} & \rightarrow & \nu/\bar{\nu} + n + 2p + ^{9}{\rm Be}\;,\label{Be9}\\
\nu/\bar{\nu} + ^{12}{\rm C} & \rightarrow & \nu/\bar{\nu} + n + p + d + ^{8}{\rm Be}\;.\label{Be8}
\end{eqnarray}
Based on our estimation, the other Atm-$\nu$ NC reactions are negligible.

In this analysis, we consider the seven reactions in Eqs.~(\ref{C11})-(\ref{Be8}) for the background analysis. The cross sections of these Atm-$\nu$ NC interactions can be obtained from the GENIE (3.0.6)~\cite{Andreopoulos:2015wxa} and NuWro (19.02) \cite{Golan:2012rfa} generators. Both the GENIE and NuWro generators use physics models, which rely in different choices for the nuclear models, the axial mass $M_A$, the underlying reaction processes, and final state interaction (FSI) models. Determining the best model among the available options presents a significant challenge currently. Therefore, we use a total of five generator physics models to estimate the event rates of the seven NC processes with the help of the Atm-$\nu$ fluxes at the JUNO site \cite{Honda:2015fha}. The details of the generator models utilized in this study are presented in Table~\ref{tab:summary_Atm_model}. For the de-excitation of the residual nuclei, we use the TALYS code  \cite{Koning:2012zqy} and take the same strategy as that of Refs.~\cite{JUNO:2022lpc, Cheng:2020aaw}. Sec.~\ref{subsec:bkg_n} will explain how these different estimates are used to predict this background.

\begin{table*}[]
    \begin{center}
    \caption{Summary of generator models (G represents GENIE~\cite{Andreopoulos:2015wxa}, N represents NuWro~\cite{Golan:2012rfa}) for atmospheric neutrino interactions. The Local Fermi Gas (LFG), Spectra Function (SF), and relativistic Fermi gas model with “Bodek-Ritchie” modifications (BRRFG) are used as the nuclear models, with $M_{A}$ representing the axial mass. The Berger-Sehgal (BS) model is employed for the coherent and diffractive production (COH) and nuclear resonance production (RES). For the two particle two hole (2p2h), the empirical model is utilized for GENIE. Final state interactions are described by the hN, a traditional hadron-nucleon intranuclear cascade model, and hA, a custom model that offers a more empirical description of the effect of multiple hadron-nucleon interactions.} 
    \vspace{0.1cm}
	\label{tab:summary_Atm_model}
        \resizebox{\textwidth}{!}{%
	\begin{tabular}{p{1.5cm}|p{1.2cm}|p{1.8cm}|p{2.7cm}|p{2.5cm}|p{2.0cm}|p{1.0cm}}
			\hline
			\hline
			\bf Models & version     & $M_{A}$[GeV] & Nuclear Model & RES \& COH & 2p2h & FSI \\
			\hline
			     G1   & 3.0.6       & 0.96         & LFG           & BS       & empirical & hN \\
			     G2    & 3.0.6       & 0.96         & LFG           & BS       & empirical & hA \\
			     G3    & 3.0.6       & 0.96         & BRRFG         & BS       & empirical & hN \\
			     N1    & 19.02       & 1.03         & SF            & default  & default   & default \\
			     N2    & 19.02       & 1.03         & LFG           & default  & default   & default \\
            \hline
            \hline
        \end{tabular}
        }
    \end{center}
\end{table*}

\section{Event selection}\label{sec:es}

To enhance JUNO's sensitivity to the neutron invisible decays, we should choose the proper event selection criteria to effectively suppress backgrounds while maintaining high signal efficiency. In this section, the muon veto strategy is first introduced before the event selection to reduce cosmogenic isotopes, radioactivity, and other backgrounds. Then, beyond the basic event selection criteria based on signal physical characteristics aimed at primarily selecting the vast majority of the neutron invisible decay events, a multiplicity cut strategy is designed to select real triple signals. 

\subsection{Muon veto strategy}\label{subsec:bkgveto}
A muon veto must be used to reduce the impact of long-lived isotopes produced by the spallation of cosmic-ray muons in the JUNO LS \cite{JUNO_YB}. Its primary goals are to maximize the rejection of $^{9}$Li/$^{8}$He and to reduce triple coincidences produced through triple isotope production by the same muon, both of which are dominant backgrounds in the search for neutron invisible decays. JUNO has developed two kinds of muon veto strategies to investigate the reactor antineutrinos \cite{JUNO:2022mxj} and solar neutrinos \cite{JUNO:2020hqc}. Unlike the single signal from solar neutrinos and the prompt-delayed signal from reactor antineutrinos, a muon veto strategy tailored for the unique triple coincidence signal expected in this analysis is developed using a 10-year MC simulation as follows, based on the corresponding physical feature distribution:
\begin{itemize}
\item  For all muons passing the water pool and/or the central detector, a veto of 3 ms after each muon is applied over the whole detector to
suppress spallation neutrons and short-lived radioactive isotopes.

\item For all muons passing the water pool and/or the central detector, a 3 m spherical volume around any spallation neutron candidates is vetoed for 30~s and 10 s in the \Ninv and \NNinv analyses, respectively.
 
\item For well-reconstructed muon tracks in the central detector caused by single or two far-apart ($>$ 3 m) muons, a veto of 2 s is applied to candidate events with reconstructed vertices smaller than 2 m away from each track.
 
\item For events containing two close and parallel muons ($< $~3~m), which constitute roughly 0.6\% of muon-related events, a single track is often reconstructed. A cylindrical veto with a radius of 3.5 m around this track is applied for 5 s. 

\item For events where a track cannot be properly reconstructed, which amount to about 2\% of all muon-related events and occur primarily when more than two muons go through the detector simultaneously, a 0.2 s veto is applied over the whole detector volume for \Ninv due to its larger $\Delta T_{23}$.

\end{itemize}
The above muon veto strategy can effectively reject the cosmogenic long-lived isotopes while keeping a high signal efficiency. For neutron invisible decay events, the dead time and dead volume introduced by the muon veto strategy may result in one of the signals in the triple signal being vetoed, leading to a decrease in the signal efficiency. The efficiencies of the four de-excitation modes in Eqs.~(\ref{M1})-(\ref{M4}) after the above muon veto are 65.7 $\pm$ 0.2(stat.)\%, 65.5 $\pm$ 0.2(stat.)\%, 80.8~$\pm$~0.2(stat.)\%, and 78.3 $\pm$ 0.2(stat.)\%, respectively. Meanwhile, the long-lived cosmogenic isotopes and neutrons have been well suppressed. After applying the above muon veto strategy, the $^9$Li and $^8$He rates  are approximately 0.07 (0.1)~/day and 0.02 (0.06)~/day for \Ninv ($nn \rightarrow inv$). Compared with 127~/day and 40~/day before the muon veto, the $^{9}$Li and $^{8}$He rates have been reduced by a factor of over 500.

\begin{table}[!h]
    \begin{center}
    \caption{Basic selection criteria used for \Ninv and $ n n \rightarrow inv$.}
	\label{tab:evt_sel_criteria}
		\vspace{1mm}
        \begin{tabular}{ m{2cm}|m{2cm}|m{2cm} }
        \hline
        \hline
        Quantity & $n \rightarrow inv$ & $nn \rightarrow inv$ \\
        \hline
        $R_{1,2,3}$  [m] & $<$ 16.7 & $<$ 16.7 \\
        \hline
        $E_{1}$ [MeV] & 0.7-12  & 0.7-30 \\
        $E_{2}$ [MeV] & 1.9-2.5 & 1.9-2.5 \\
        $E_{3}$ [MeV] & 1.5-3.5 & 3.0-16.0 \\
        \hline
        $\Delta T_{12}$ [ms] & $<$ 1 & $<$ 1 \\
        $\Delta T_{23}$ [s] & 0.002-100 & 0.002-3.0 \\
        \hline
        $\Delta R_{12}$ [m] & $<$ 1.5 & $<$ 1.5 \\
        $\Delta R_{23}$ [m] & $<$ 1.5 & $<$ 1.5 \\
        $\Delta R_{13}$ [m] & $<$ 1.0 & $<$ 1.0  \\
        \hline
        \hline
        \end{tabular}
    \end{center}
\end{table}

\subsection{Event selection}\label{subsec:selection criteria}

As previously mentioned, a fiducial volume cut of $R_{i}<$ 16.7 m is chosen for both \Ninv and $ n n \rightarrow inv$ in order to reduce the contributions of external radioactivity and fast neutrons to backgrounds. As shown in Fig.~\ref{fig:radioactivity_fv}, the fiducial radius cut can reject most external radioactivity events. According to the signal characteristics as demonstrated in Sec.~\ref{subsec:sig_features}, we set the basic event selection criteria for $E_{i}$, $\Delta R_{ij}$, and $\Delta T_{ij}$ as listed in Table~\ref{tab:evt_sel_criteria}. For the \Ninv and \NNinv analyses, the prompt energy $E_1$ is restricted to the range of $[0.7, 12]$~MeV and $[0.7, 30]$ MeV, respectively. For the delayed signal from the neutron capture, we require the reconstructed energy $E_2 \in$~$[1.9, 2.5]$ MeV and the time interval of $\Delta T_{12} <$ 1.0 ms. The decay energy $E_3$ from the daughter nuclei in Eqs.~(\ref{M1})-(\ref{M4}) takes the range of $[1.5, 3.5]$ MeV ($[3.0, 16.0]$ MeV) in the search of \Ninv ($ nn \rightarrow inv$). The corresponding time interval $\Delta T_{23}$ lies in the range $[0.002, 100]$~s ($[0.002, 3.0]$~s). To further reduce the accidental coincidence backgrounds, the surviving triple signals are restricted to occur in proximity to each other: $\Delta R_{12}<$ 1.5 m, $\Delta R_{23}<$ 1.5 m, and $\Delta R_{13}<$ 1.0 m. As shown in Fig.~\ref{fig:Singal_characteristic}, these basic selection criteria can select the vast majority of the neutron invisible decay events.

\begin{table*}[h!]
	\centering
	\caption{Summary of selection efficiencies (\%) of four de-excitation modes for \Ninv and $ n n \rightarrow inv$.} 
		\label{tab:selection_eff}
		\vspace{0.5mm}
		\small
            \resizebox{\textwidth}{!}{%
		\begin{tabular}{l|c|c|c|c}
			\hline
			\bf Selection Criterion  &  \multicolumn{2}{|c|}{\Ninv} & \multicolumn{2}{|c}{\NNinv} \\
			\cline{2 - 5}
			\bf                     & \Na               & \Nb               & \NNa                      & \NNb                      \\
			\hline
			All triple signals      & 100               & 100               & 100                       & 100                       \\
            Muon Veto               & 65.7 $\pm$ 0.2    & 65.5 $\pm$ 0.2    & 80.8 $\pm$ 0.2            & 78.3 $\pm$ 0.2            \\
			Fiducial Volume         & 83.5 $\pm$ 0.4    & 82.7 $\pm$ 0.4    & 82.9 $\pm$ 0.4            & 83.1 $\pm$ 0.4            \\
			Event Selection         & 75.4 $\pm$ 0.9    & 89.7 $\pm$ 0.3    & 89.2 $\pm$ 0.3            & 83.5 $\pm$ 0.3            \\
            Multiplicity Cut        & 93.8 $\pm$ 0.1    & 93.8 $\pm$ 0.1    & 99.9 $\pm$ $\cal{O}$$(10^{-4})$   & 99.9 $\pm$ $\cal{O}$$(10^{-4})$   \\
            \hline
            \bf Combined Selection  & 38.8 $\pm$ 0.5    & 45.6 $\pm$ 0.3    & 59.7 $\pm$ 0.4            & 54.3 $\pm$ 0.4            \\
			\hline
		\end{tabular}
	}	
\end{table*}

Applying the above selection criteria, we calculate the signal selection efficiencies of the four de-excitation modes in Eqs.~(\ref{M1})-(\ref{M4}) as listed in Table~\ref{tab:selection_eff}. The efficiency of the fiducial volume cut is about 83\%, which is consistent with $(16.7/17.7)^3 \approx 84.0\%$. The selection criteria of $E_i$, $\Delta T_{ij}$, and $\Delta R_{ij}$ preserve most of the signals. For $^{11}{\rm C^*} \rightarrow  n +  ^{10}{\rm C}$, the relatively small efficiency of 75.4\% is due to the energy threshold of $E_1 >$ 0.7~MeV, which can reduce some signals, as shown in the top-left panel of Fig.~\ref{fig:Singal_characteristic}. For $^{10}{\rm C^*} \rightarrow   n + p +  ^{8}{\rm B}$, the requirement of $\Delta T_{23} < 3.0$~s rejects many $^{8}$B decay signals because of its half-life time of 0.77~s. 
In Table~\ref{tab:selection_eff}, we present a summary of event selection efficiencies. According to Ref.~\cite{JUNO:2022mxj}, the systematic uncertainties are from the fiducial volume cut (2 cm vertex bias) and the event selection. 

Most of the de-excitation events from Eqs.~(\ref{M1})-(\ref{M4}) will generate a triple coincidence signal in the JUNO LS. In reality, many single events from the natural radioactivity and cosmogenic isotopes can appear between the delayed and decay signals of neutron invisible decays, owing to the large time interval $\Delta T_{23}$. In this case, the triple coincidence signal can easily form a quadruple or higher coincidence with an uncorrelated single event. To correctly select the triple coincidence signals from the experimental data and reject the influence of single events, this study uses two kinds of multiplicity cut methods. Method I first selects the prompt-delayed signal, which should satisfy the selection criteria of $E_1$, $E_2$, $\Delta R_{12}$, and $\Delta T_{12}$.
In addition, we have the following requirements:
\begin{itemize}
	\item No trigger with $0.7 < E < 12$ MeV in a 1 ms window before the prompt signal;
	\item No other events during $\Delta T_{12}$; 
	 \item No trigger with $0.7 < E < 12$ MeV in a 1 ms window before/after the decay signal;
    \item Only one prompt-delayed signal in a 100 s window before the decay signal. 
\end{itemize}
Method II firstly searches for the neutron capture signal. Its requirements are the following:
\begin{itemize}
	\item Only 1 satisfied trigger falls in [-1, 0] ms;
     \item No other neutron capture signals within a spherical volume with a radius of 3 m during [-3.001, 3.001] s;
    \item Only 1 satisfied trigger falls in [0, 3] s.
\end{itemize}
The biggest difference is that Method I has a more strict space limit than Method II. It requires the absence of candidate signals that meet the criteria anywhere within the detector space, whereas Method II only requires no candidate signals within a specific selected spatial range. Hence, Method I gives a slightly lower multiplicity cut efficiency of $93.8 \pm 0.1$ ($97.9 \pm$ $\cal{O}$ $(10^{-2})$) compared to the $95.7 \pm 0.1$ ($99.9 \pm$ $\cal{O}$ $(10^{-4})$) of Method II for \Ninv ($ n n \rightarrow inv$). 
In the following parts of this work, we conservatively take Method I to analyze \Ninv due to its larger time interval of $\Delta T_{23}$. Method II is applied to $nn \rightarrow inv$.

Based on Table~\ref{tab:selection_eff} and  the detection efficiencies discussed in Sec.~\ref{subsec:sig_features}, we calculate the signal efficiencies of the four de-excitation modes in Eqs.~(\ref{M1})-(\ref{M4}) as listed in Table~\ref{tab:bkg_summary}. The signal efficiencies $\epsilon_{n1}$ and $\epsilon_{n2}$ correspond to the de-excitation modes in Eqs.~(\ref{M1}) and (\ref{M2}), while the signal efficiencies $\epsilon_{nn1}$ and $\epsilon_{nn2}$ correspond to those of Eqs.~(\ref{M3}) and~(\ref{M4}).

\section{Background estimation}\label{sec:bkg_estimation}

We have introduced all possible background sources in Sec.~\ref{subsec:background_sources}. Most backgrounds are formed by the accidental coincidence of two background sources, while the Atm-$\nu$ NC interactions can directly produce a triple coincidence signal. In this section, we estimate all types of backgrounds for \Ninv and \NNinv based on the event selection criteria described in Sec.~\ref{sec:es}. Subsequently, the pulse shape discrimination technique and the multivariate analysis method will be employed to further suppress the background. For instance, for the IBD events followed by an uncorrelated single event, the prompt signals from neutron invisible decays and IBD events are generated by neutrons and positrons, respectively, which exhibit different pulse shapes \cite{psd_measurement, Mollenberg:2014pwa}. Consequently, the pulse shape discrimination (PSD) technique can be employed to reject background events. Furthermore, the correlations among $E_{1}$, $E_2$, $E_{3}$, $\Delta R_{12}$, $\Delta R_{23}$, $\Delta R_{13}$, $\Delta T_{12}$, and $\Delta T_{23}$ exhibit variations between signal and accidental coincidence background. The discrepancies among the time, spatial and energy distributions suggest that the employment of the multivariate analysis (MVA) method will be instrumental in effectively discerning the signals from the backgrounds. To ensure the reliability of the results, we have employed two distinct background estimation methods and suppression techniques during the analysis that cross-validated each other. The background estimation utilizes both the MC approach and computational methods. For suppression methods, one approach does not integrate the PSD into the MVA, whereas the other incorporates the PSD as a variable along with the basic features. It is worth noting that these methodologies are employed for background analysis. In the ensuing sections, we have opted to elaborate on the method that exhibited very slightly superior performance for each decay mode.

\subsection{ \Ninv analysis}\label{subsec:bkg_n}

To simplify, we refer to the prompt-delayed signal as a Double signal. Meanwhile, the events with no other correlated triggers are referred to as ``Single". We first estimate the Double+Single background rates ($R_{\rm Double+Single}$), originating from the coincidence of a doubly correlated event (IBD, $^{9}$Li/$^{8}$He, $^{13}$C($\alpha$, n)$^{16}$O, fast neutron) with Single.
\begin{equation}\label{eq:double_singles_n}
    R_{\rm Double +Single} = R_{\rm Double}(1 - e^{-R_{\rm Single} \cdot P_{(\Delta R_{23,13})} \cdot \Delta T_{23}})  ,
\end{equation}
where $R_{\rm Double}$ is the rate of Double signals after applying the selection criteria of $E_{1}$, $E_{2}$, $\Delta R_{12}$, $\Delta T_{12}$, the fiducial volume cut, and the muon veto. $R_{\rm Single} = 0.71$ Hz represents the Single rate with the energy $E_3$ in the range of [1.5, 3.5]~MeV, including $0.64$ Hz of radioactivity and $0.07$ Hz of isotopes. $P_{(\Delta R_{23,13})}$ is the survival probability of backgrounds after the spatial cuts $\Delta R_{23}$ and $\Delta R_{13}$.

\begin{figure*}[]
	\centering
	\includegraphics[width=1\textwidth]{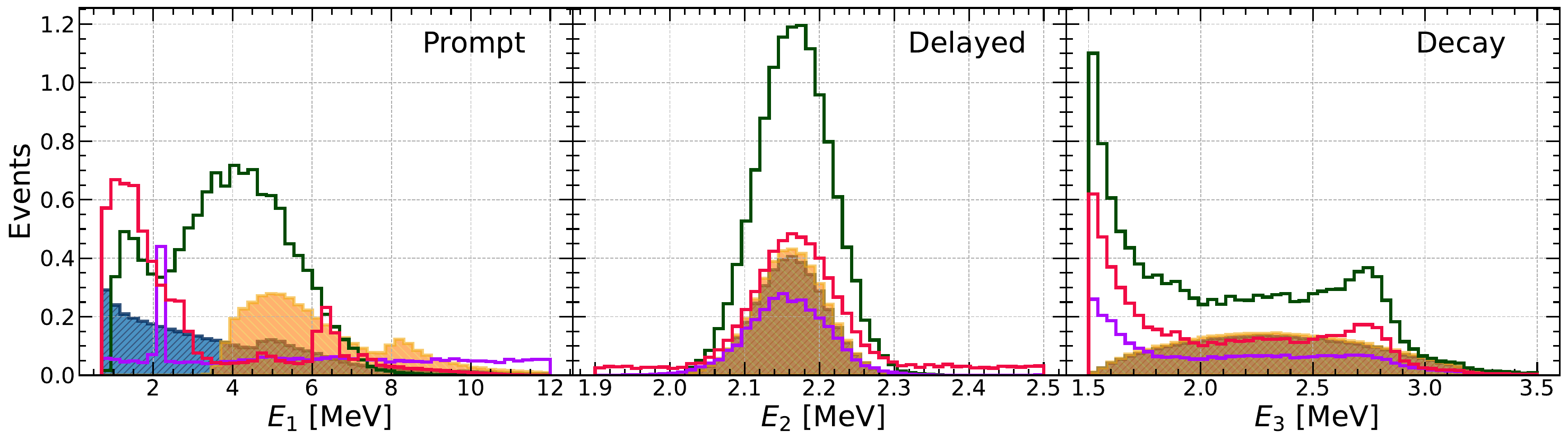}
	\includegraphics[width=1\textwidth]{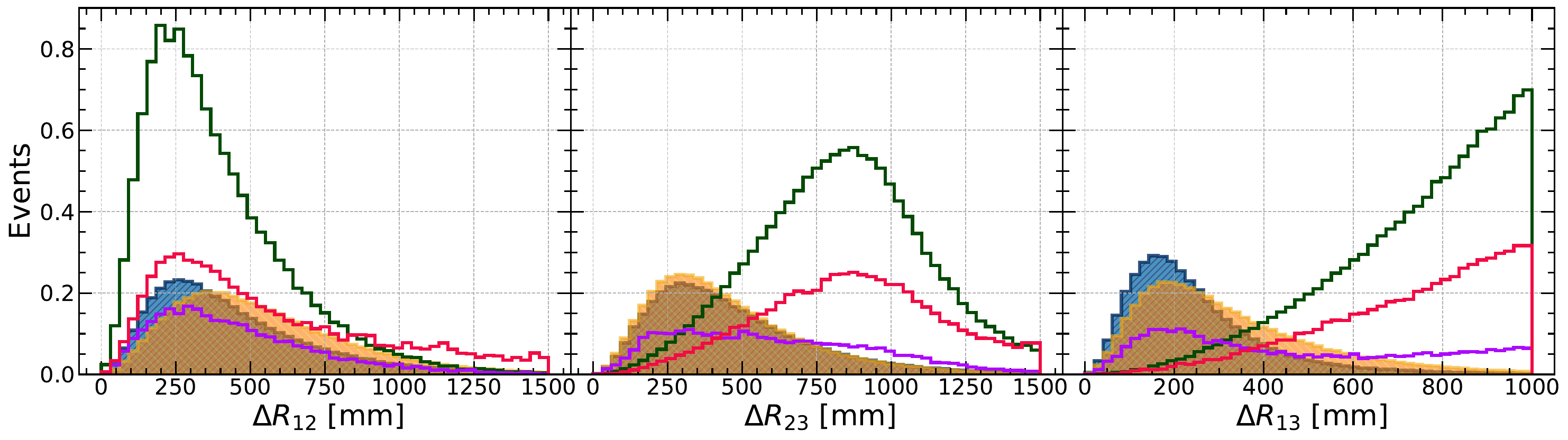}
	\vspace{-0.14cm}
	\includegraphics[width=1\textwidth]{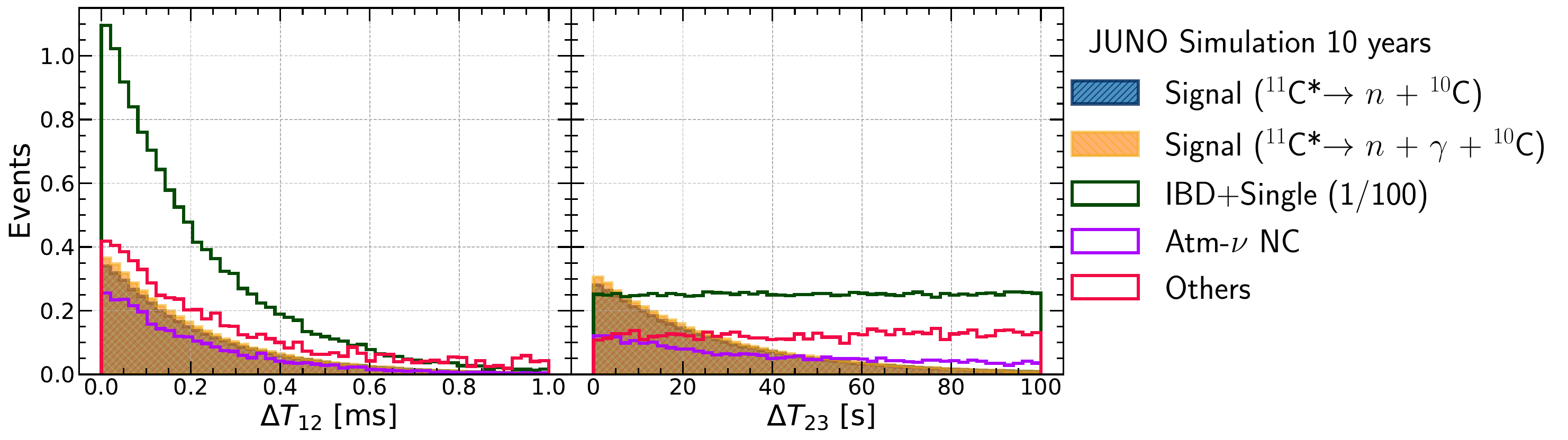}
	\vspace{-0.14cm}
    \caption{The $E_1$, $E_2$, $E_3$, $\Delta R_{12}$ $\Delta R_{13}$, $\Delta R_{23}$, $\Delta T_{12}$, and $\Delta T_{23}$  distributions of \Ninv and  backgrounds. The vertical axis represents the event number over a period of 10 years, and x-axis for each plot corresponds to the selection criterion in Table~\ref{tab:evt_sel_criteria}. The expected signal number is calculated based on the JUNO sensitivity of this article, namely $\tau/B( n \rightarrow inv) = 5.0 \times 10^{31}$ yr. The number in parentheses in the legend represents the scale factor.}
	\label{fig:sig_distibution_n}
\end{figure*}

After 10 years, the expected rates for this type of background, based on the spatial distribution of Single, are summarized in Table~\ref{tab:bkg_summary}. Note that $^9$Li/$^8$He+Single rates account for the possibility of both the correlated $^9$Li/$^8$He events and the single event originating from the same muon shower. In Fig.~\ref{fig:sig_distibution_n}, we show the energy, time interval, and spatial distributions of the IBD+Single background. As a comparison, we also plot the distributions of $n \rightarrow inv$, which is calculated based on the event selection in Sec.~\ref{sec:es} and the sensitivity reported in Sec.~\ref{sec:sen}, namely $\tau/B( n \rightarrow inv) = 5.0 \times 10^{31}$ yr. Using a similar method, the accidental triple coincidence from three Single has been estimated to be 1.46 $\pm$ 0.05~/10~years. 

The systematic uncertainty of the IBD+Single rate arises from the uncertainties of both the IBD rate and the Single rate. Here we adopt a relative systematic uncertainty of 2\% for the world reactor IBD rate based on Ref.~\cite{JUNO:2022mxj}. For the Single rate, it consists of contributions from natural radioactivity and cosmogenic isotopes. The systematic uncertainty of the radioactivity rate is expected to be small, as we can measure the rate precisely after JUNO starts taking data. Therefore, a conservative uncertainty of 1\% is used for the accidental rate~\cite{JUNO:2022mxj}. For the cosmogenic isotopes rate, we assign a larger uncertainty of 20\% according to the KamLAND experiment \cite{KamLAND:2009zwo}. For the $^{13}$C($\alpha$, n)$^{16}$O, an uncertainty of 50\%  from Ref.~\cite{JUNO_YB} is applied. 

\begin{figure}[!t]
	\centering
	\includegraphics[width=0.48\textwidth]{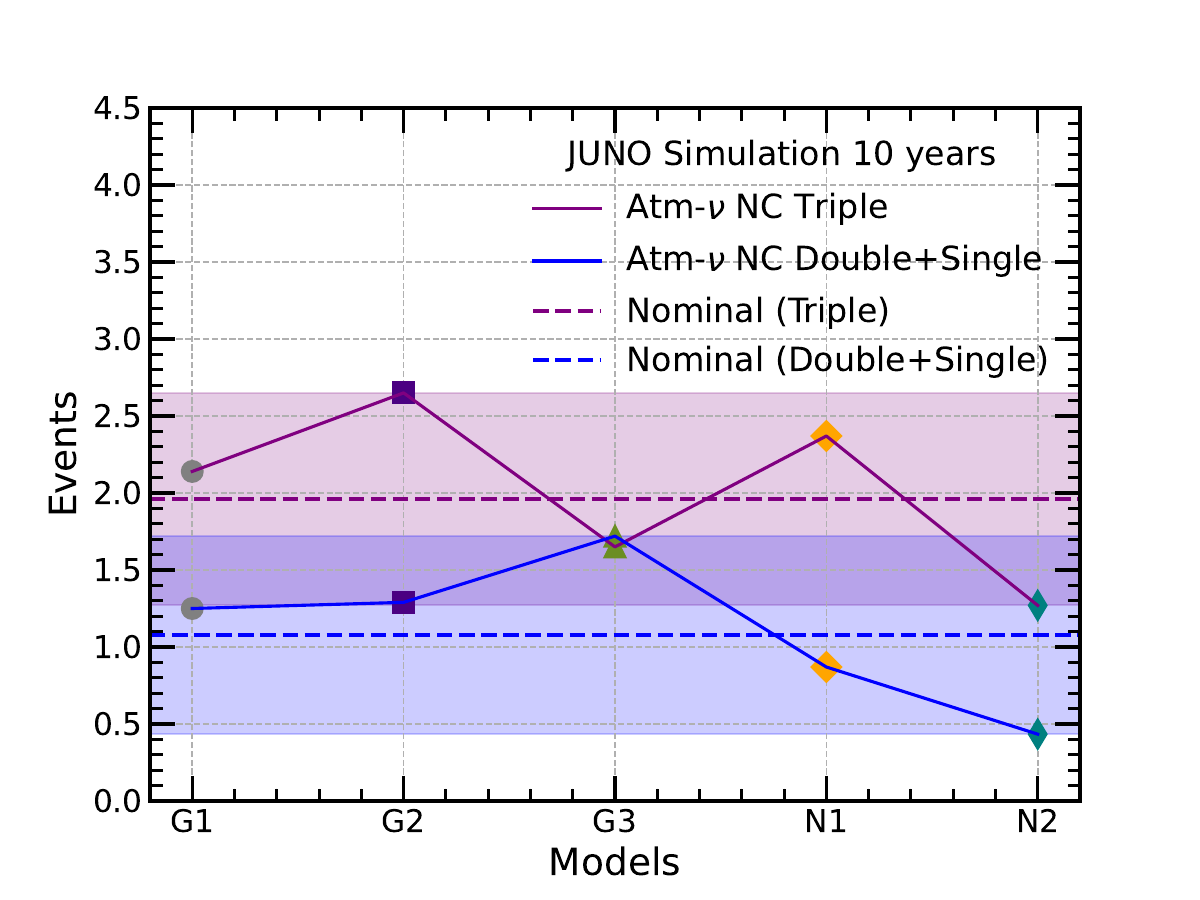}
	\includegraphics[width=0.48\textwidth]{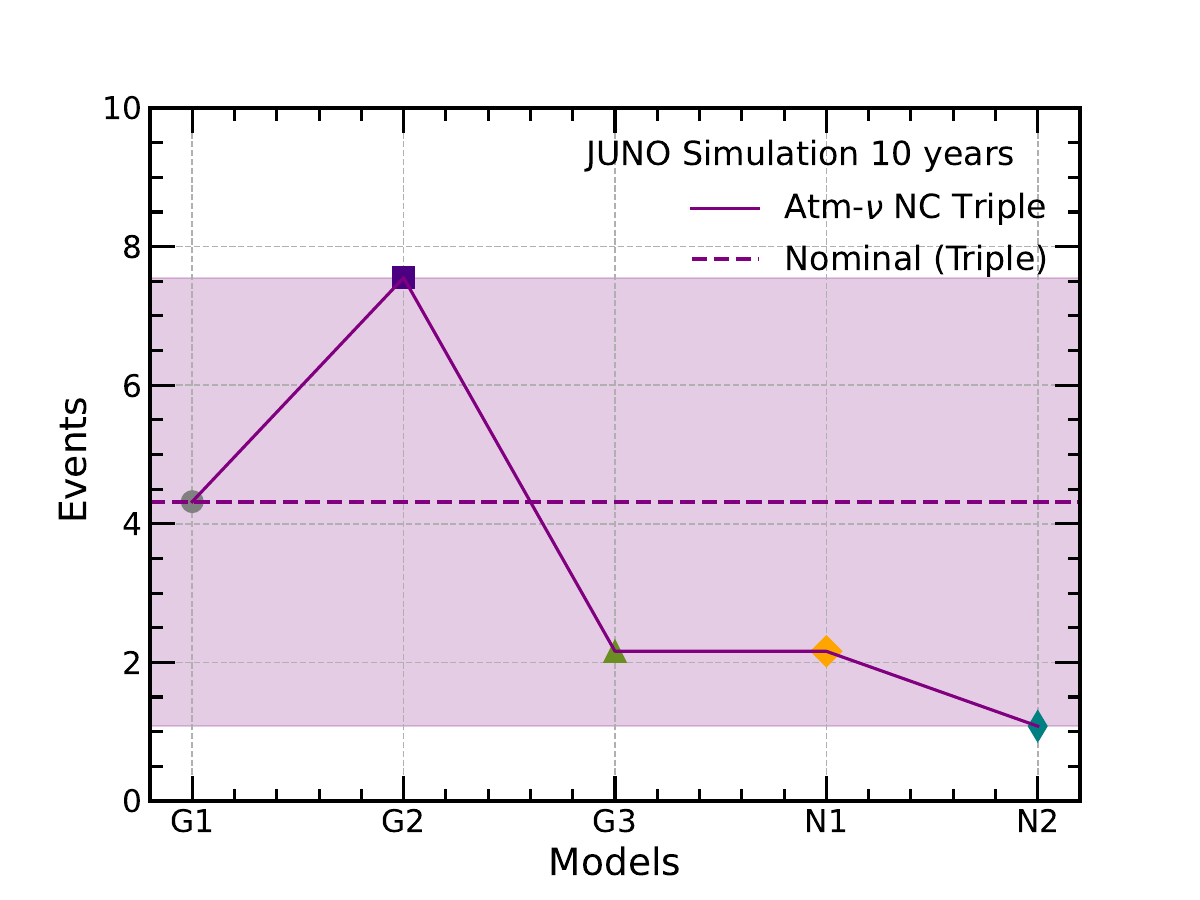}
	\caption{\label{fig:NC}The Atm-$\nu$ NC background rates from five typical generator models in Table~\ref{tab:summary_Atm_model}  after basic event selection criteria for \Ninv(left) and \NNinv(right). The horizontal axis describes different models, while the vertical axis represents the background rate per 10 years. The dashed line within the corresponding color rectangular region denotes the average of the maximum and minimum among five models, which we take as the nominal background rate. Its systematic uncertainty is chosen as half of the difference between the maximal and minimal values. }
\end{figure}

For the Atm-$\nu$ NC background, we first calculate the total event rate of three interactions in Eqs.~(\ref{C11})-(\ref{Li8}). The five typical generator models in Table~\ref{tab:summary_Atm_model} give significantly different predictions, as shown in Fig.~\ref{fig:NC}. The rate derived from the NuWro SF nuclear model (N2) is lower compared to other models. The averaged rate from the three GENIE models is higher than that from the two NuWro models. To reasonably estimate the Atm-$\nu$ NC background, we select the maximal and minimal values among the five predicted event rates, and then take their mean value as the nominal event rate. The systematic uncertainty from the interaction cross section is chosen as half of the difference between the maximum and minimum values. This uncertainty estimation method can fully encompass all five models. Using Eq.~(\ref{eq:double_singles_n}), we analyze the accidental coincident background between the Atm-$\nu$ NC reactions in Eqs.~(\ref{B10})-(\ref{Be8}) and Single. All five models indicate that the Atm-$\nu$ NC Double+Single background is not negligible for $n \rightarrow inv$, as shown in Fig.~\ref{fig:NC}. We finally consider the statistical error and the systematic uncertainty from the Atm-$\nu$ flux. Previous research has already delineated the systematic uncertainty estimation from the Atm-$\nu$ flux at the JUNO site \cite{Cheng:2020aaw}. Given that this research also employs the identical Honda Flux with an energy range of 0.1-20 GeV, we conservatively apply a systematic uncertainty of 30\% to the flux. After considering all uncertainties, the final total Atm-$\nu$ NC background number is 3.0 $\pm$ 1.1 in 10 years, as listed in Table~\ref{tab:bkg_summary}. Their energy, time interval, and spatial distributions have been shown in Fig.~\ref{fig:sig_distibution_n}. In the $E_1$ spectrum of NC, a peak is observed from the $2 n +^{10}$C reaction channel. Our simulation observed a case where some final state nucleons of the NC interaction process from GENIE had zero momentum in GENIE (3.0.6), potentially affecting the prompt signal composition for $2 n+^{10}$C. Additional calculations were performed to address this issue. Our estimations with 5 models and 2 generators (GENIE and NuWro) show that its impact on the final results is negligible.

\begin{figure}[!t]
	\centering
	\includegraphics[width=0.48\textwidth]{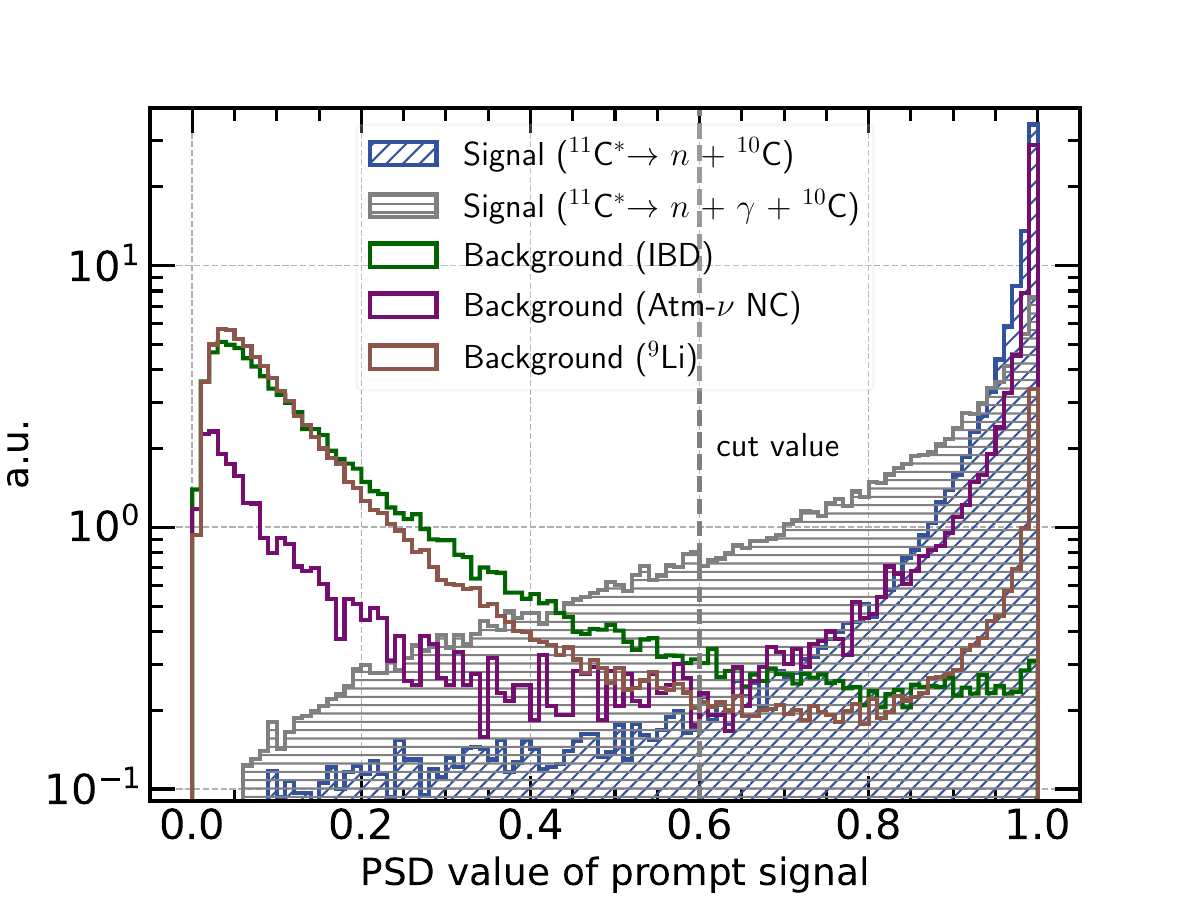}
	\includegraphics[width=0.48\textwidth]{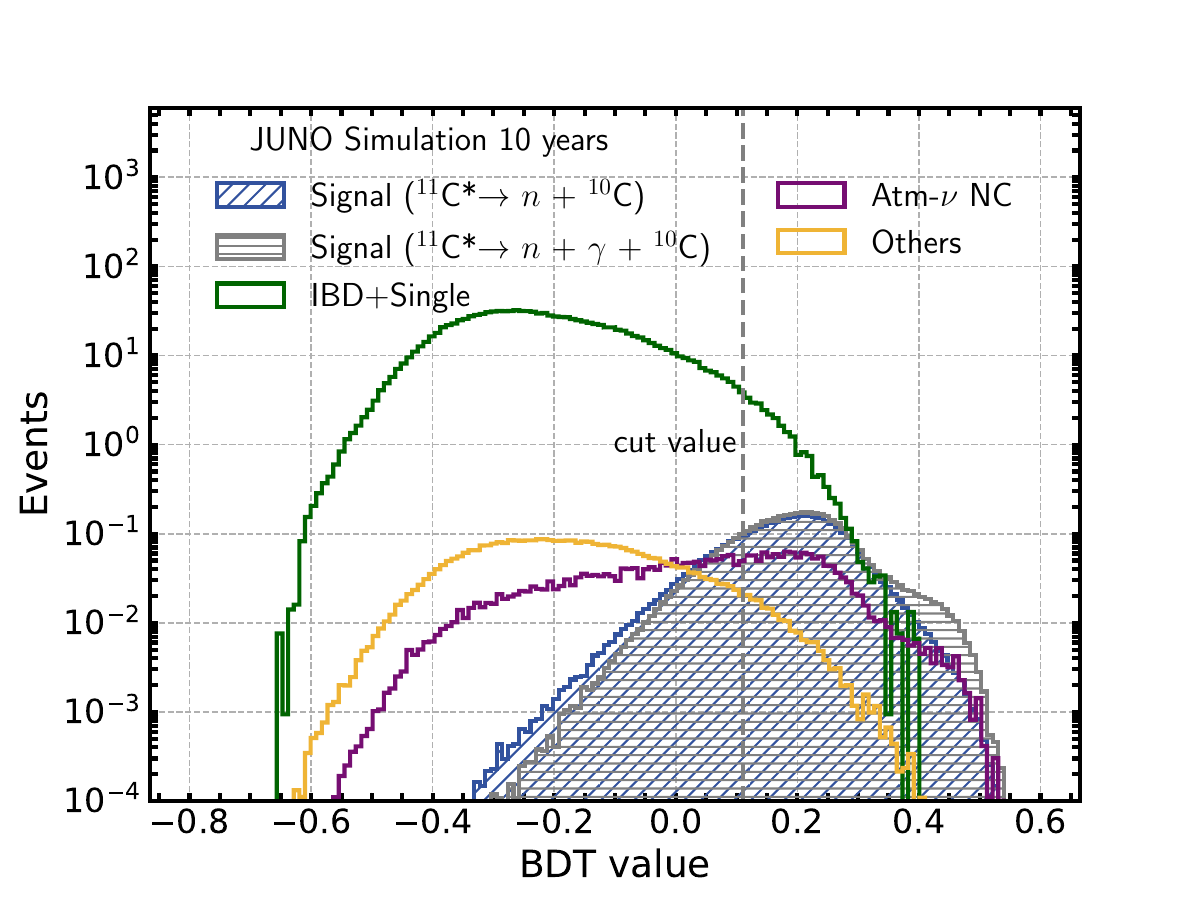}
	\caption{\label{fig:ML_n}The PSD (left) and MVA (right) separation capabilities for $ n \rightarrow inv$. The left plot is normalized to unity to highlight the performance of the PSD, while right plot shows the BDT performance and the vertical axis represents the event number over a period of 10 years.
	 }
\end{figure}

After the basic event selections, many backgrounds still remain, as listed in Table~\ref{tab:bkg_summary}, especially from IBD+Single. The prompt signal for the vast majority of background events is caused by a positron. However, the neutron invisible decays generate the prompt signal through the neutron elastic and inelastic scattering processes as described in Sec.~\ref{subsec:sig_features}. Note that the deposited energies of positrons and energetic neutrons in the LS have distinct photon emission time profiles \cite{psd_measurement, Mollenberg:2014pwa}. Based on this feature, JUNO is planning to use the PSD technique described in~\cite{JUNO:2022lpc,Cheng:2023zds} to separate the IBD events from the Atm-$\nu$ NC events. Here we primarily utilize the same PSD tool to distinguish the prompt signal of the neutron invisible decays and two kinds of dominant backgrounds (IBD+Single, $^{9}$Li+Single). As shown in the left panel of Fig.~\ref{fig:ML_n}, the PSD technique can effectively identify the invisible decay signals from the two backgrounds, especially from the IBD+Single. There is a tail around a PSD value of 1.0 for $^{9}$Li events, which can be attributed to the fact that the prompt signal of $^{9}$Li contains the kinetic energy of both the $\beta ^{-} $ and neutron. Note that the Atm-$\nu$ NC background cannot be suppressed effectively since its prompt signal also comes from the energetic neutrons.         

In addition to the PSD technique, some significant differences between the signals and backgrounds can be clearly observed in Fig.~\ref{fig:sig_distibution_n}. These differences among the time, space, and energy distributions imply that the multivariate analysis method can be powerful in further separating the signal from the background. Here we employ the Boosted Decision Trees (BDT) and consider eight input variables, including $E_{1}$, $E_2$, $E_{3}$, $\Delta R_{12}$, $\Delta R_{23}$, $\Delta R_{13}$, $\Delta T_{12}$, and $\Delta T_{23}$. The right panel of Fig.~\ref{fig:ML_n} shows the BDT output distributions. It is clear that the MVA method can reject most backgrounds except for the Atm-$\nu$ NC triple events. Applying the PSD and MVA methods, the residual backgrounds and the signal efficiency are derived as tabulated in Table~\ref{tab:bkg_summary}. The cut values for the PSD and BDT values have been chosen as 0.60 and 0.11 for the best sensitivity, respectively. The chosen values correspond to the BDT and PSD at the optimum $\epsilon_{sig}/S_{90}$, where $S_{90}$ represents the average upper limit of the signal number at a 90$\%$ confidence level. Further details of the optimization can be found in the appendix. The IBD+Single background is significantly suppressed relative to the Atm-$\nu$ NC background. This is because of the inherent dissimilarity in PSD values between the IBD events and the signals. The IBD+Single and Atm-$\nu$ NC events contribute predominantly to the backgrounds after PSD+MVA suppression, with the other backgrounds contributing negligibly.

\subsection{ \NNinv analysis}\label{subsec:bkg_nn}

To ensure the reliability of the results, we have developed two independent approaches to estimate the backgrounds for both \Ninv and $ n n \rightarrow inv$. In contrast with the approach described in Sec.~\ref{subsec:bkg_n} for \Ninv analysis, we employed a numerical calculation approach to estimate the background of $ n n \rightarrow inv$. For the Double+Single background, we first divide the fiducial sphere of radius 16.7 m into 4658 concentric shells of equal volume due to the non-uniform radial distribution of external radioactivity, and assume the single events are uniformly distributed in each shell. Secondly, due to the selection criteria of $\Delta R_{12} < 1.5$ m and $\Delta R_{23} < 1.5$ m, a spherical volume with a radius of 1.5 m is chosen around the delayed signal. For a prompt-delayed event in the $i^{th}$ shell, we can calculate the fraction $f_i^j$ of the volume of the $j^{th}$ shell inside this sphere to the total volume of this shell. Finally, we use the following formula to calculate the background rate of the Double+Single events:
\begin{equation}\label{eq:IBD_singles_nn}
     R_{\rm Double+Single}  = \sum_{i=1} R^{i}_{\rm Double}(1 - e^{-\sum_{j} f_i^j \cdot R^{j}_{\rm Single} \Delta T_{23}}) P_{\Delta R_{13}},
\end{equation}
where $R^{i}_{\rm Double}$ is the prompt-delayed event rate in the $i^{th}$ shell after applying the selection criteria of $E_{1}$, $E_{2}$, $\Delta R_{12}$, $\Delta T_{12}$, the fiducial volume cut, and the muon veto strategy. Similarly, $R^{j}_{\rm Single}$ denotes the Single rate in the $j^{th}$ shell following the $3.0 \,{\rm MeV} < E_{3} < 16.0$~MeV selection. $P_{\Delta R_{13}}$ is the survival probability of backgrounds after the $\Delta R_{13}$ cut.

\begin{figure*}[]
	\centering
	\includegraphics[width=1.\textwidth]{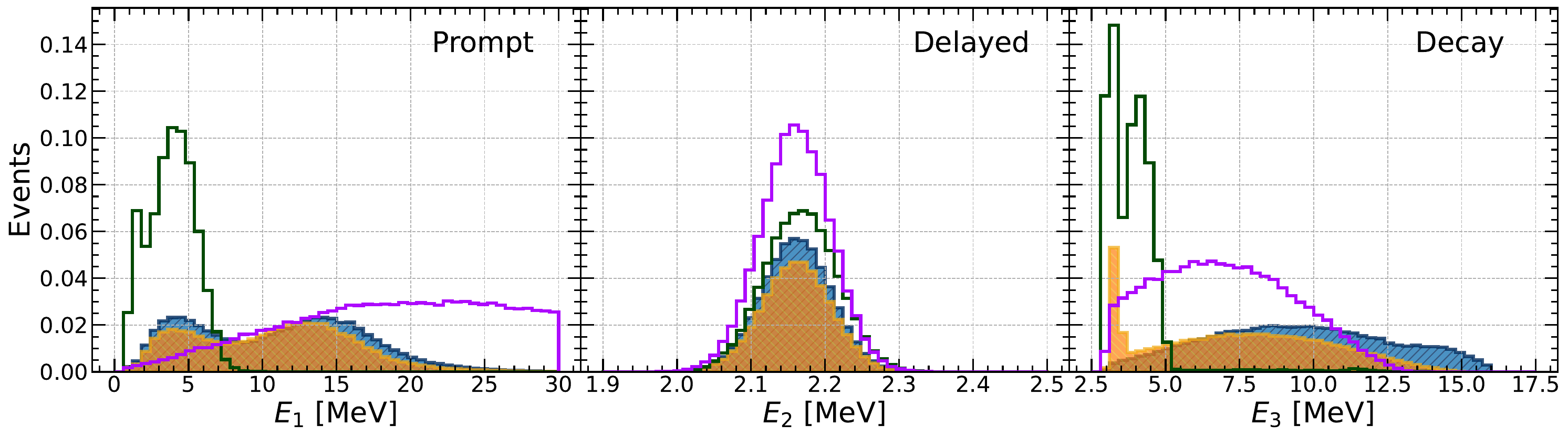}
	\includegraphics[width=1.\textwidth]{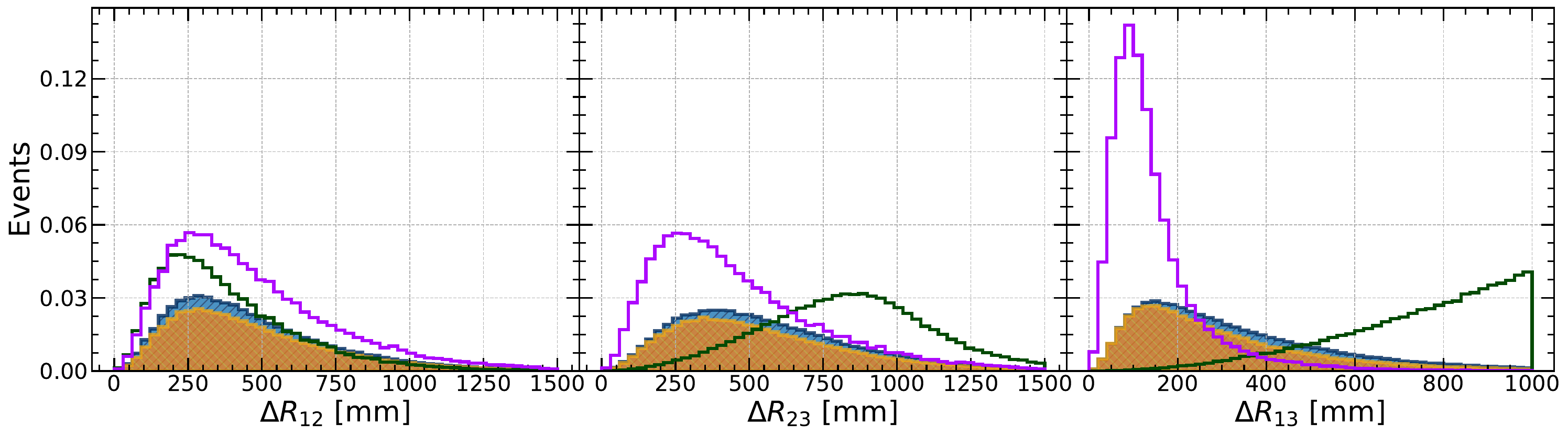}
 	\vspace{-0.14cm}
	\includegraphics[width=1.\textwidth]{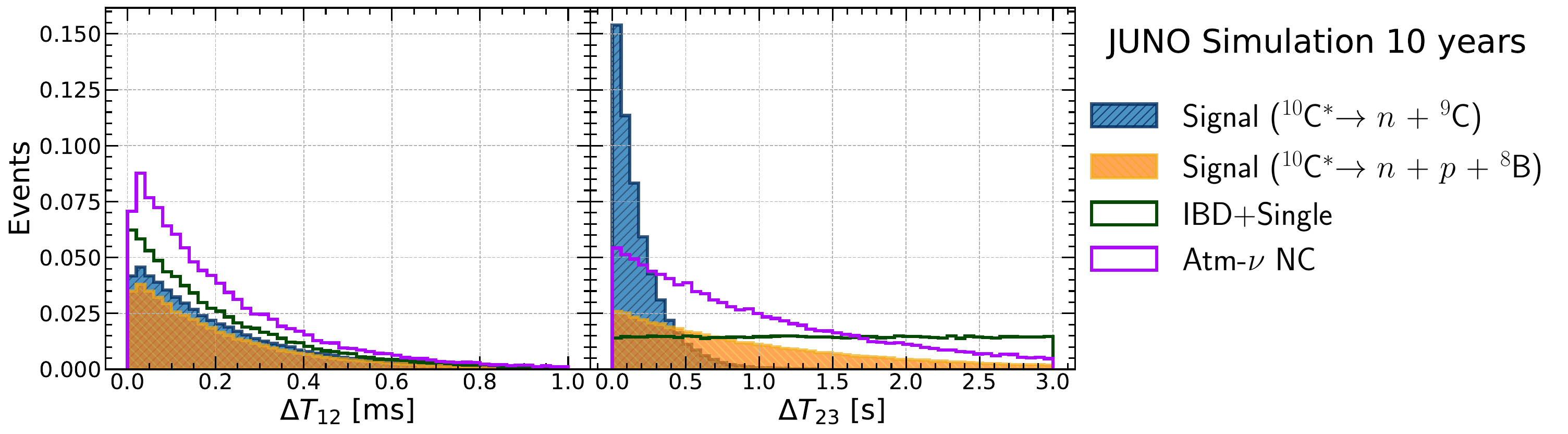}
	\caption{The $E_1$, $E_2$, $E_3$, $\Delta R_{12}$, $\Delta R_{13}$, $\Delta R_{23}$, $\Delta T_{12}$, and $\Delta T_{23}$  distributions of \NNinv and two dominant backgrounds. The vertical axis represents the event number over a period of 10 years, and x-axis for each plot corresponds to the selection criterion in Table~\ref{tab:evt_sel_criteria}. The expected signal number is calculated based on the JUNO sensitivity of this article, namely $\tau/B(n n \rightarrow inv) = 1.4 \times 10^{32}$ yr.}
	\label{fig:sig_distibution_nn}
\end{figure*}

By utilizing the event rates of $R^{i}_{\rm Double}$ and $R^{j}_{\rm Single}$ in each shell, we estimate the background rates from the Double+Single. It is found that the expected number of IBD+Single event is 3.01 $\pm$ 0.09 in 10 years. In Fig.~\ref{fig:sig_distibution_nn}, we plot the energy, time interval, and spatial distributions of the IBD+Single background, where the signal distributions are calculated based on the event selection in Sec.~\ref{sec:es} and the JUNO sensitivity of $\tau/B( n \rightarrow inv) = 1.4 \times 10^{32}$ yr in this work. Note that other Double+Single backgrounds from $^{9}$Li/$^{8}$He, $^{13}$C($\alpha$, n)$^{16}$O and fast neutrons, and the accidental triple coincidence of three Single are negligible. Besides the accidental coincidence with Single, $^{9}$Li/$^{8}$He can also form a triple coincidence event with an isotope from the same muon shower. Based on the 10 years of simulation data of cosmic muons, we find that the final rate of this kind of background is 0.13 $\pm$ 0.13 per 10 years. A relative statistical uncertainty of 100\% has been assigned.

For $ n n \rightarrow inv$, one can easily find that the Atm-$\nu$ NC Double+Single background rate (0.1/10 years) will be negligible compared to the IBD+Single rate. Consequently, we focus exclusively on the Atm-$\nu$ NC triple coincident events in Eqs.~(\ref{C11})-(\ref{Li8}). In fact, the dominant channel is $ \nu/\bar{\nu} \,+\, ^{12}$C$ \, \rightarrow  \nu/\bar{\nu} + 3p + n +\,^{8}$Li. This is because the selection criteria of $3.0$ MeV $ < E_3 <16.0$ MeV can reject all $^{11}$C and most of $^{10}$C. In addition, $\Delta T_{23} < 3.0$ s can also remove many $^{10}$C since its half-life is 19.29 s. Here, we still employ five generator models to estimate the rate of Atm-$\nu$ NC triple events, as shown in Fig.~\ref{fig:NC}. Using the same uncertainty estimation method in Sec.~\ref{subsec:bkg_n}, the Atm-$\nu$ NC background rate is given by 4.3 $\pm$ 3.5 in 10 years, including both the cross section and flux uncertainties. The energy, time interval, and spatial distributions of the Atm-$\nu$ NC background have also been illustrated in Fig.~\ref{fig:sig_distibution_nn}.

\begin{figure}[!t]
	\centering
	\includegraphics[width=0.48\textwidth]{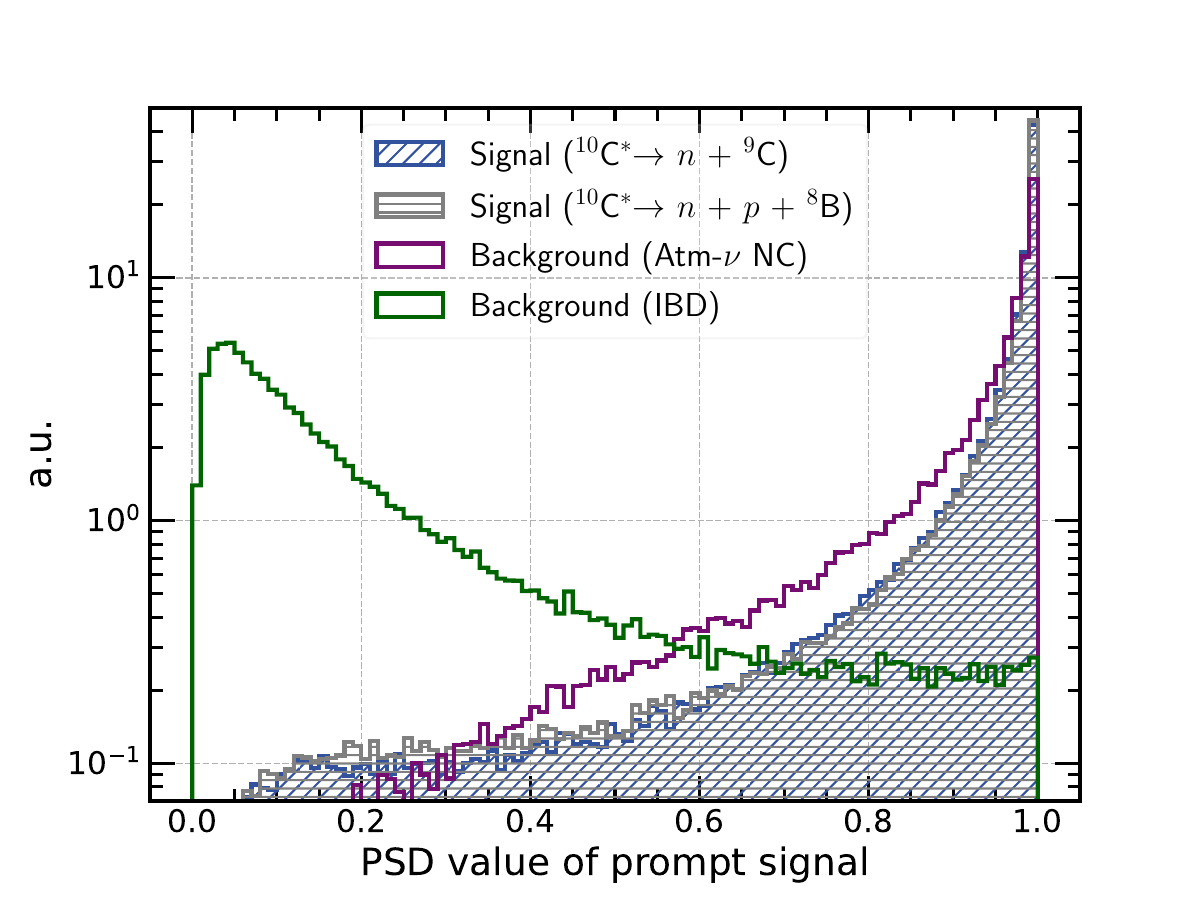}
	\includegraphics[width=0.48\textwidth]{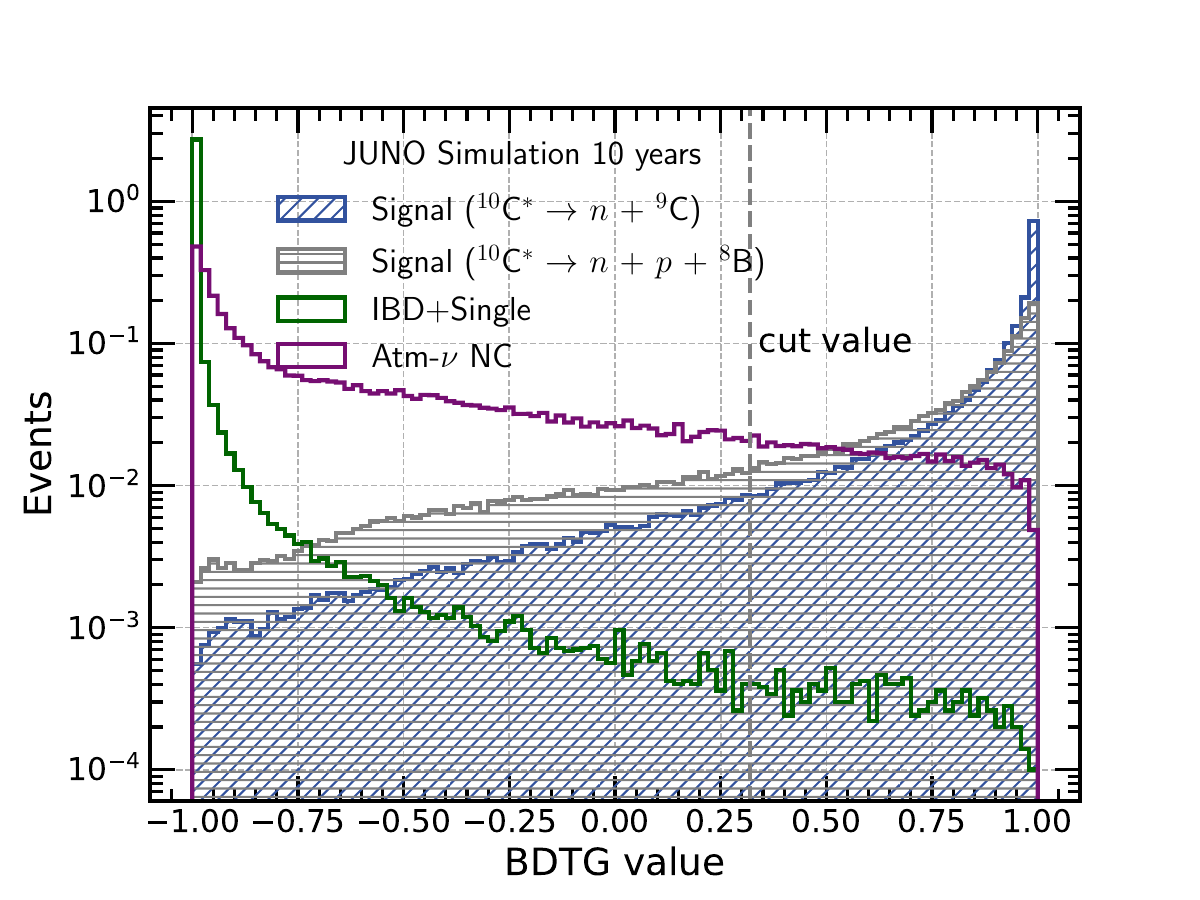}
	\caption{\label{fig:ML_nn} The PSD (left) and MVA (right) separation capabilities for $n n \rightarrow inv$.  The left plot is normalized to unity to highlight the performance of the PSD, while right plot show the BDTG performance and the vertical axis represents the event number over a period of 10 years.}
\end{figure}

After the basic event selections, the estimated total background rate is 7.4 $\pm$ 3.5 per 10 years, as listed in Table~\ref{tab:bkg_summary}. Here, we further suppress these backgrounds by using the PSD and MVA methods. As shown in the left panel of Fig.~\ref{fig:ML_nn}, the PSD technique can also effectively distinguish the invisible decay signals from the IBD+Single background. To enhance the reliability of our results, the \NNinv analysis incorporates some differences compared to $ n \rightarrow inv$, such as the incorporation of the PSD value into the MVA and the use of a BDT with Gradient Boosting (BDTG). We take the PSD output, along with other basic features, as an input variable for the MVA training. Whereas PSD and MVA are treated as two independent variables for the \Ninv analysis. In the right panel of Fig.~\ref{fig:ML_nn}, we plot the MVA output distribution for $n n \rightarrow inv$. It indicates that this MVA method has a good capability to distinguish the signal from the IBD+Single and Atm-$\nu$ NC backgrounds. In the following analyses, the BDTG cut value of 0.32 is chosen to maximize the sensitivity. Further optimization details are provided in the appendix. In this case, the residual background rates and signal efficiencies after PSD+MVA have been listed in Table~\ref{tab:bkg_summary}. The Atm-$\nu$ NC events and the $^{9}$Li/$^{8}$He+Single become the dominant backgrounds after PSD+MVA suppression. It is worth emphasizing that the IBD+Single background is suppressed to a negligible level.  

\begin{table*}[]
    \begin{center}
    \caption{Summary of the background event rates (10 years) and signal efficiencies before and after the PSD+MVA for \Ninv and $ n n \rightarrow inv$.  }
    \vspace{0.2cm}
	\label{tab:bkg_summary}
        \resizebox{\textwidth}{!}{%
		\begin{tabular}{c|c|c|c|c}
			\hline
			\hline
			\bf Backgrounds  &  \multicolumn{2}{|c|}{\Ninv} & \multicolumn{2}{|c}{\NNinv} \\
			\cline{2 - 5}
			\bf (10 years) & Basic selection & PSD + MVA & Basic selection & PSD + MVA \\
        \hline 
        IBD + Single                           & 1235 $\pm\,$50    & 2.72 $\pm\,$ 0.10     & 3.01 $\pm\,$ 0.09 & 0.0110 $\pm\,$ 0.0003     \\
        \hline 
        Atm-$\nu$ NC                            & 3.0 $\pm\,$ 1.1   & 0.89 $\pm\,$ 0.67     & 4.3 $\pm\,$ 3.5   & 0.55 $\pm\,$ 0.63         \\
        \hline
        $^{13}$C($\alpha$,n)$^{16}$O + Single  & 3.4 $\pm\,$ 1.4   & 0.036 $\pm\,$ 0.013   & --                & --                        \\
        \hline
        $^{9}$Li/$^{8}$He + Single             & 1.55 $\pm\,$ 0.39 & 0.29 $\pm\,$ 0.17     & 0.13 $\pm\,$ 0.13 & 0.13 $\pm\,$ 0.13         \\    
        \hline
        Accidental                              & 1.46 $\pm\,$ 0.05 & 0.095 $\pm\,$ 0.004   & --                & --                        \\
        \hline
        Total & 1244 $\pm$ 50  & 4.03 $\pm$ 0.70& 7.4 $\pm$ 3.5 & 0.69 $\pm$ 0.64 \\
        \hline
 		\bf Signal efficiency  &  \multicolumn{2}{|c|}{\Ninv} & \multicolumn{2}{|c}{\NNinv} \\
		\cline{2 - 5}
		\bf (\%)   & Basic selection & PSD + MVA & Basic selection & PSD + MVA \\
        \hline 
         $\epsilon_{n(nn)1}$ & 35.6 $\pm\,$ 0.2 & 23.5 $\pm$ 0.2 & 54.0 $\pm\,$ 0.3 & 48.2 $\pm\,$ 0.3  \\
        \hline
         $\epsilon_{n(nn)2}$ & 43.6 $\pm\,$ 0.3 & 30.3 $\pm$ 0.3 & 49.2 $\pm\,$ 0.3 & 36.3 $\pm\,$ 0.3 \\
        \hline
        \hline
        \end{tabular}
        }
    \end{center}
\end{table*}

\section{Sensitivity}\label{sec:sen}

Based on the MC simulation, the background rates, signal efficiencies, and their uncertainties have been derived as listed in Table~\ref{tab:bkg_summary}. The JUNO sensitivity for \Ninv can be calculated as follows:
\begin{eqnarray}
\tau/B( n \rightarrow { inv} ) > N_0 \, T \, \sum_{i=1,2}\epsilon_{ni} \, B_{ni}
/ S_{90} \;, \label{eq:tau_limit}
\end{eqnarray}
where $B(n \rightarrow { inv} )$ represents the branching ratio of the invisible decay mode $n \rightarrow {inv}$ when one neutron undergoes the baryon number-violating decay. $N_0 = 1.76 \times 10^{33}$ is the number of $s$-shell neutrons, and $T$ is the JUNO running time. The signal efficiencies $\epsilon_{ni}$ can be found in Table \ref{tab:bkg_summary}. $B_{n(nn)i}$ are the de-excitation branching ratios for single (two) neutron invisible decay, which can be known from Eqs.~(\ref{M1})-(\ref{M4}). A systematic uncertainty of 30\% will be considered, stemming from the theoretical prediction of de-excitation branching ratios of highly excited residual nuclei \cite{Kamyshkov:2002wp}. $S_{90}$ is the upper limit of the detected signal number at a 90\% confidence level (C.L.). It depends on the number of observed events and the background level. For $nn \rightarrow inv$, the calculation of sensitivity follows a similar procedure but with corresponding $s$-shell $nn$ pairs number of $N_0 = 8.8 \times 10^{32}$, signal efficiencies $\epsilon_{nni}$ and branching ratios $B_{nni}$.

\begin{figure}[!t]
	\centering
	\includegraphics[width=0.7\textwidth]{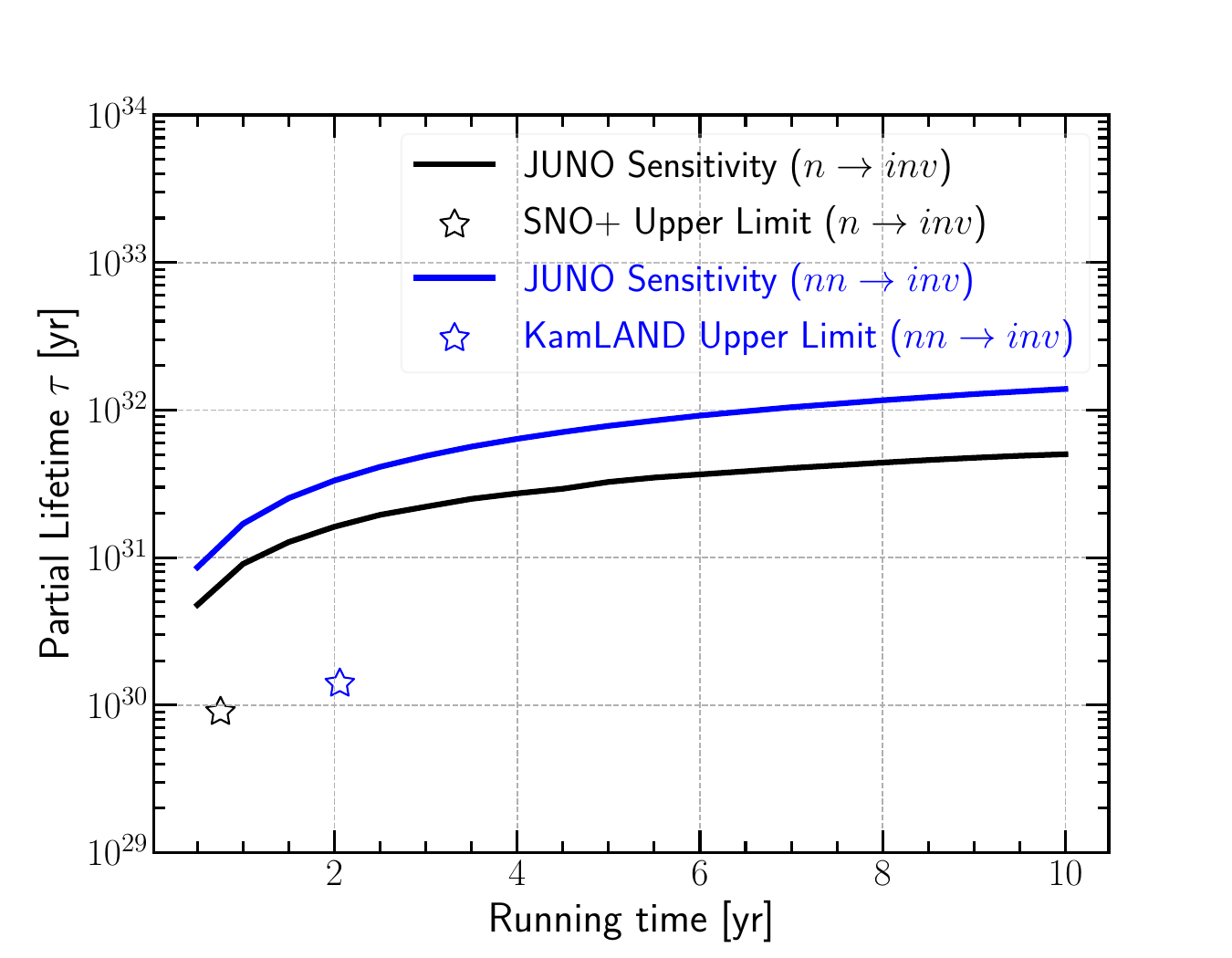}
    \caption{JUNO sensitivities to \Ninv and \NNinv as a function of the running time at 90\% C.L.. SNO+ and KamLAND give the current best upper limits in the search of \Ninv and \NNinv based on experimental data, respectively.}
	\label{fig:Tau_vs_running_time}
\end{figure}

The likelihood contours method \cite{Feldman:1997qc} (usually denoted as the Feldman-Cousins method) is employed to calculate $S_{90}$ for \Ninv and $n n \rightarrow inv$. Here we report the sensitivities using the average upper limits, which are obtained by the following formula \cite{Gomez-Cadenas:2010zcc}:
\begin{equation}\label{eq:sensitivity_formular}
    S_{90} = \sum_{n=0}^{\infty} \, P(n|b) \, {U}(n|b) ,
\end{equation}
where ${U} (n|b)$ is a function yielding the upper limit for a given observation $n$ and a predicted background $b$ through the likelihood contours method in 90\% confidence level. $P(n|b)$ is the Poisson  distribution for the pure background. The systematic uncertainties in signal efficiency and backgrounds are accounted for by integrating over probability density functions that parametrize these uncertainties~\cite{Conrad:2002kn}. The large background uncertainty in \NNinv may lead to a negative value when assuming a Gaussian distribution. Thus, we choose the Log-Normal distribution~\cite{Conrad:2002kn} to describe the uncertainties of the background rate. The signal efficiencies, background rates, and their uncertainties, as shown in Table~\ref{tab:bkg_summary}, result in the following JUNO sensitivities to \Ninv and \NNinv after 10 years of data taking at the 90\% C.L.:
\begin{equation}\label{eq:sensitivity}
    \begin{split}
    \tau/B( n \rightarrow { inv} )  > 5.0\times 10^{31} \, {\rm years}, \\
    \tau/B( nn \rightarrow { inv} ) > 1.4 \times 10^{32} \, {\rm years}.
    \end{split}
\end{equation}
The predicted sensitivity of \Ninv($ n n \rightarrow inv$) is almost one(two) orders of magnitude better than the SNO+ (KamLAND) results. In Fig.~\ref{fig:Tau_vs_running_time}, we plot the JUNO sensitivity as a function of the running time. It is found that with two years of data, JUNO will give an order of magnitude improvement compared with the current best limits: $\tau/B( n \rightarrow { inv} )> 9.0 \times 10^{29}$ yr and $\tau/B( nn \rightarrow { inv} ) > 1.4 \times 10^{30}$ yr. 

To ensure the reliability of the results in Eq.~(\ref{eq:sensitivity}), we use the analysis method in Sec.~\ref{subsec:bkg_n} (Sec.~\ref{subsec:bkg_nn}) and the corresponding multiplicity cut strategy to study \NNinv ($ n \rightarrow inv$). It is found that different analysis methods only slightly influence the final sensitivities. In addition, we have also varied the MVA cuts to investigate the sensitivity changes. The predicted sensitivities are not significantly affected by the MVA cut value when it falls within the range of [0, 0.5]. To assess the impact of Atm-$\nu$ NC as the main background on the predicted sensitivities, we enlarge their nominal values and uncertainties. We observe that for both decay modes, escalating the Atm-$\nu$ NC uncertainty to 150\% marginally affects the predicted sensitivities. Given that Atm-$\nu$ NC is one of the main backgrounds for both analyses, the sensitivity is expected to change by about 10\% and 20\% when doubling the nominal value of the Atm-$\nu$ NC event rate, respectively.

\section{Conclusion}\label{sec:conclusion}

In conclusion, we have investigated neutron invisible decays in the JUNO LS detector. The triple coincidence characteristic arising from the invisible decays of $s$-shell neutrons in $^{12}{\rm C}$ has been briefly described, providing insight into the development of dedicated selection criteria applied to the signal sample generated with the full simulation. To correctly select triple coincidence signals from the experimental data and minimize the influence of the uncorrelated single signals, we have developed two muon veto strategies and implemented two types of multiplicity cut methods tailored to the characteristics of both the signal and backgrounds. On the other hand, we have conducted a detailed estimation of all potential backgrounds, which have been classified into six categories: IBD+Single, Isotope($^{9}$Li/$^{8}$He)+Single, FN+Single, $^{13}$C($\alpha$,n)$^{16}$O+Single, Accidental backgrounds, and Atm-$\nu$ NC events. It is observed that the IBD+Single and Atm-$\nu$ NC events are the dominant backgrounds. To suppress backgrounds further, we employ pulse shape discrimination and multivariate analysis techniques in both searches. After 10 years of JUNO data taking, the expected background numbers for \Ninv and \NNinv are 4.07~$\pm$~0.68 and 0.69 $\pm$ 0.64, with final signal efficiencies of 26.7\% and 42.3\%, respectively. Finally, we have found that with two years of data, JUNO will yield an improvement by an order of magnitude with respect to the current best limits. After 10 years of data taking, the expected sensitivities for JUNO are $\tau/B( n \rightarrow { inv} ) > 5.0 \times 10^{31} \, {\rm yr}$ and $\tau/B( nn \rightarrow { inv} ) > 1.4  \times 10^{32} \, {\rm yr}$ at the 90\% confidence level.

\section*{Acknowledgement}
We are grateful for the ongoing cooperation from the China General Nuclear Power Group.
This work was supported by
the Chinese Academy of Sciences,
the National Key R\&D Program of China,
the CAS Center for Excellence in Particle Physics,
Wuyi University,
and the Tsung-Dao Lee Institute of Shanghai Jiao Tong University in China,
the Institut National de Physique Nucl\'eaire et de Physique de Particules (IN2P3) in France,
the Istituto Nazionale di Fisica Nucleare (INFN) in Italy,
the Italian-Chinese collaborative research program MAECI-NSFC,
the Fond de la Recherche Scientifique (F.R.S-FNRS) and FWO under the ``Excellence of Science – EOS” in Belgium,
the Conselho Nacional de Desenvolvimento Cient\'ifico e Tecnol\`ogico in Brazil,
the Agencia Nacional de Investigacion y Desarrollo in Chile,
the Charles University Research Centre and the Ministry of Education, Youth, and Sports in Czech Republic,
the Deutsche Forschungsgemeinschaft (DFG), the Helmholtz Association, and the Cluster of Excellence PRISMA+ in Germany,
the Joint Institute of Nuclear Research (JINR) and Lomonosov Moscow State University in Russia,
the joint Russian Science Foundation (RSF) and National Natural Science Foundation of China (NSFC) research program,
the MOST and MOE in Taiwan,
the Chulalongkorn University and Suranaree University of Technology in Thailand,
and the University of California at Irvine in USA. 

\appendix
\section*{Appendix: Optimization of sensitivity}
We separately studied the $\epsilon_{sig}/S_{90}$ variation trends with the BDT/PSD and BDTG cut values for two invisible decay modes, as shown in the following Fig.~\ref{fig:sensitivity_vs_MVA_value}. The left panel depicts the case of \Ninv, where a two-dimensional scan was performed using PSD and BDT values, with the white star indicating the location of the optimal scan value. For \NNinv, as PSD is used as an input variable for BDTG, it can be represented by a one-dimensional curve. Similarly, the red star represents the optimal scan point. It should be noted that, in order to simplify the calculation during the optimization process, the systematic uncertainties of signal efficiency and background event counts were not taken into account.
\begin{figure}[!htbp]
    \raisebox{-4mm}
    {
        \begin{minipage}[t]{0.54\textwidth}
        \centering
        \includegraphics[width=\textwidth]{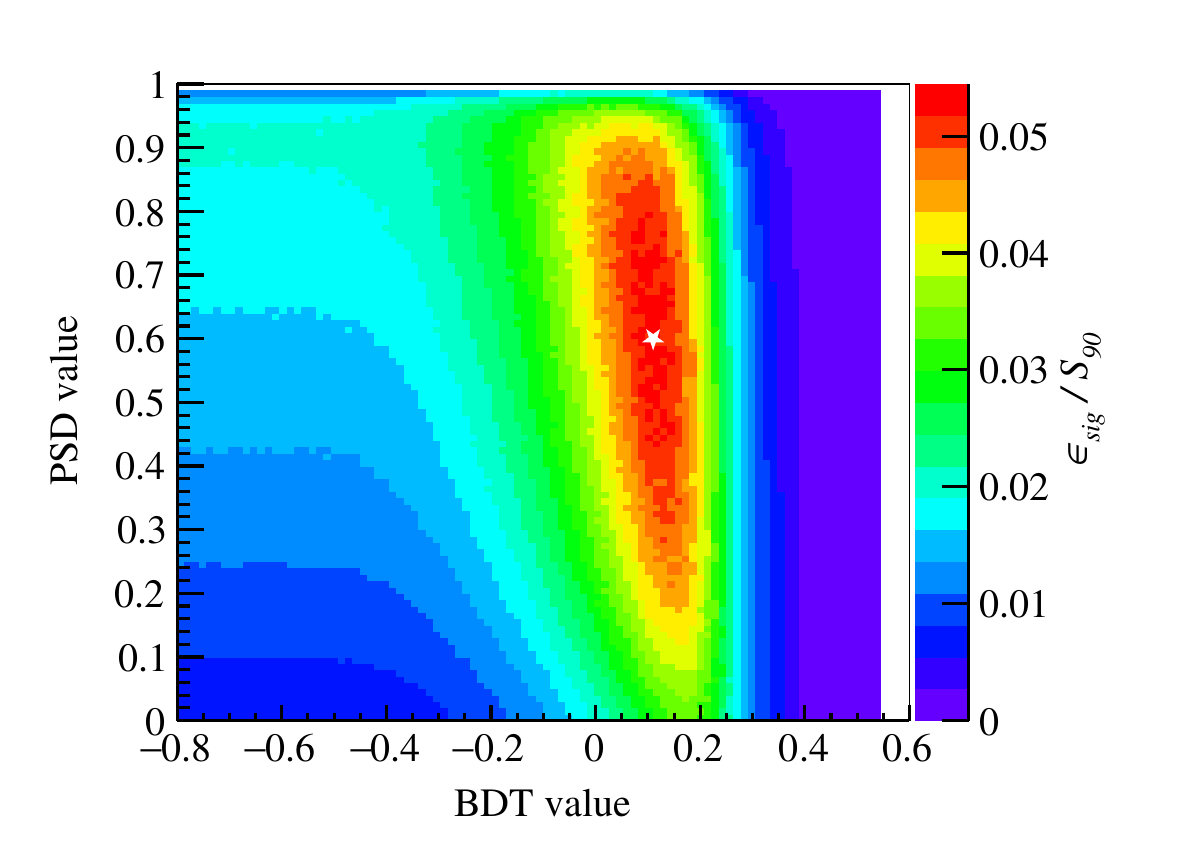}
        \end{minipage}
    }
    \begin{minipage}[t]{0.5\textwidth}
        \centering
        \includegraphics[width=\textwidth]{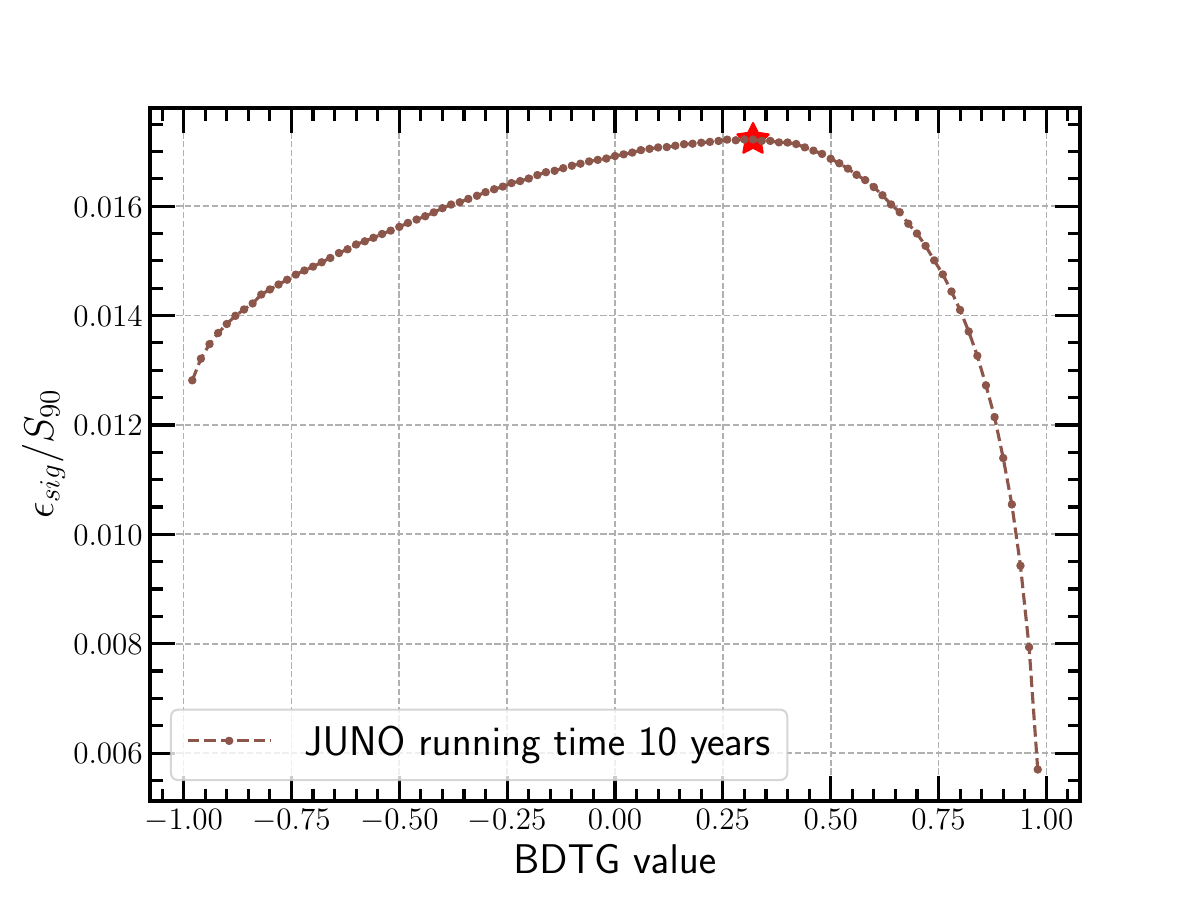}
    \end{minipage}
    \caption{ The sensitivity (without including systematic uncertainty) variation trends with the BDT/PSD cuts of \Ninv (Left) and the BDTG cut of \NNinv (Right). The white star and red star in the graph represnt the optimal value for \Ninv and \NNinv, respectively.}
    
	\label{fig:sensitivity_vs_MVA_value}
\end{figure}

\bibliographystyle{JHEP}
\bibliography{biblio.bib}

\end{document}